# Dislocation-induced stress in polycrystalline materials: mesoscopic simulations in the dislocation density formalism

D.V. Berkov, N.L. Gorn

*General Numerics Research Lab e.V.,
Moritz-von-Rohr-Str. 1A, D-07745 Jena, Germany*

## Abstract

In this paper we present a simple and effective numerical method which allows a fast FFT-based evaluation of stress generated by dislocations with arbitrary directions and Burgers vectors if the (site-dependent) dislocation density is known. Our method allows the evaluation of the dislocation stress using rectangular grid with shape-anisotropic discretization cells without employing higher multipole moments of the dislocation interaction coefficients. Using the proposed method, we first simulate the stress created by relatively simple non-homogeneous distributions of vertical edge and so called 'mixed' dislocations in a disk-shaped sample, what is necessary to understand the dislocation behaviour in more complicated systems. The main part of our research is devoted to the stress distribution in polycrystalline layers with the dislocation density rapidly varying with the distance to the layer bottom. Considering GaN as a typical example of such systems, we investigate dislocation-induced stress for edge and mixed dislocations, having random orientations of Burgers vectors among crystal grains. We show that the rapid decay of the dislocation density leads to many highly non-trivial features of the stress distributions in such layers and study in detail the dependence of these features on the average grain size. Finally we develop an analytical approach which allows to predict the evolution of the stress variance with the grain size and compare analytical predictions with numerical results.

## 1. Introduction

Dislocation-induced stress plays a very important role in many fundamental physical phenomena in mono- and polycrystals and strongly affects crystal properties relevant for various technological applications of bulk crystals and thin crystalline films (see, e.g., [Anderson2017, Hull2011] and references therein). For example, this stress can lead to undesired deformations [Yamane2016] of semiconductor layers during their growth process. Further, it plays an important role in formation and propagation of cracks [Indenbom2012]. Elastic deformation caused by this stress can result in substantial inhomogeneities of optical properties of crystals (see [Chu2014] and Ref. therein), what may be especially important in crystals intended as lenses materials for high-resolution lithography applications [Wagner2010]. Another important example is the movement of dislocations due to the stress induced by other dislocations, what can significantly change the spatial variation of many crystal properties like the carrier lifetime in semiconductors (see e.g. [Clayes2011]).

For this reason analytical and numerical methods for computation of the dislocation-induced stress for the given dislocation configuration have been intensively developed, going back to the first classical results concerning the stress field of a single infinitely long straight dislocations both of the edge and screw types ([Anderson2017], Chap. 3). Introduction of the Nye tensor [Nye1953, Kröner1958] and its generalization to the dislocation density tensor [Sandfeld2011, Cajic2011] has allowed to describe dislocation dynamics in fairly complicated systems, obeying the restriction of the dislocation line continuity and more sophisticated continuity relations of the dislocation dynamics.

In recent decades, the impressive work has been accomplished in order to achieve the desired progress in evaluating the stress field in dislocation systems treated as collections of separated dislocations. Analytical expressions for the stress tensor induced by a single dislocation segment having an arbitrary orientation and Burgers vector have been derived [Devincre1992, Arsenlis2007] and reliable regularization method for the singular expression of a dislocation stress have been suggested [Cai2006]. This has led to a successful development of algorithms and software packages for simulations of the so called discrete dislocation dynamics initially for 2D systems [Amodeo1990, Barts1995, Wang1995], then for dislocation loops [Lesar2002] and later for arbitrary 3D dislocation networks [Bulatov2006, Arsenlis2007, Kubin2013].

In most experimental situations, however, the exact configuration of dislocation lines is not known. In the most common case only the measured dislocation density is available, and the corresponding density does not obey the standard continuity restrictions posed on ideal dislocation lines (see above), e.g., due to the presence of internal defects like precipitates or crystal grain boundaries in polycrystalline materials. In this case it is desirable to have a simple and effective method for the computation of the dislocation-induced stress based only on the mesoscopic dislocation density and the information of the dominant dislocation line direction and typical Burgers vectors. In this paper we present such a method, which is able to handle dislocation systems with arbitrary site-dependent dislocation density of dislocations having any line directions and orientations of Burgers vectors. The numerical implementation of our method uses the fast Fourier transformation (FFT) technique and is thus able to handle systems with very large number of discretization cells. Further subdivision of shape-anisotropic discretization cells for the computation of the elastic stress in nearest neighboring cells allows to avoid the usage of higher multipole moments in our simulations.

The paper is organized as follows. In Sec. 2, we derive basic formulas which will be used throughout the paper to evaluate the components of the stress tensor from the known dislocation density for vertical dislocations and dislocations with an arbitrary direction. In the same section we explain our numerical implementation of these basic expressions, which allows to achieve a high accuracy of the stress evaluation using (where necessary) discretization cells with a large shape anisotropy. Sec. 3 is devoted to the test example, where the stress pattern for a spatially non-homogeneous distribution of dislocation all having the same type is analyzed. In Sec. 4 we introduce an algorithm for the stress evaluation if periodic boundary conditions (PBC) should be applied at least in two directions (a common case for the simulation of a thin layer). Explanation of the relaxation method for the calculation of the stress inside a thin layer taking into account proper elastic boundary conditions on its free surfaces and PBC in the layer plane is given in Sec. 5. Physical results and their discussion for the specific system highly important for many applications - GaN layer with the dislocation density rapidly decreasing with the distance from the layer bottom (a standard situation for several GaN growth techniques) - are presented in detail in Sec. 6. In addition to numerical results (subsections 6.1 - 6.4) we present here also an analytical approach for the evaluation of the standard deviation of stress components in dependence on the average grain size for polycrystalline films (subsections 6.5) and compare our analytical predictions with numerical results. Finally, we summarize our main findings in Sec. 7.

## 2. Dislocation-induced stress in the dislocation density formalism: basic expressions

### 2.1. General integral forms for components of the dislocation induced stress.

To derive the expression suitable for the numerical evaluation of the dislocation-induced elastic stress in the formalism of the dislocation density, we start with the expression for the components $\sigma_{\alpha\beta}^{\text{seg}}(\Delta l, \mathbf{b}, \mathbf{t})$ of the stress tensor $\hat{\sigma}^{\text{seg}}$, generated by a single segment of the dislocation with the length $\Delta l$, the Burgers vector $\mathbf{b}$ and the unit vector along the dislocation line $\mathbf{t}$.

In frames of the linear elasticity theory $\sigma_{\alpha\beta}^{\text{seg}}(\Delta l, \mathbf{b}, \mathbf{t})$ is proportional to the Burgers vector length $b = |\mathbf{b}|$. Further, if we are interested in stresses on the distance $\Delta r$ much larger than the segment length $\Delta l$ ($\Delta r \gg \Delta l$), this stress is also proportional to $\Delta l$. Hence we can write the following general expression for the stress from a single dislocation segment:

$$\sigma_{\alpha\beta}^{\text{seg}}(\Delta l, \mathbf{b}, \mathbf{t}; \Delta \mathbf{r}) = \Delta l \cdot b \cdot f_{\alpha\beta}(\mathbf{e_b}, \mathbf{t}; \Delta \mathbf{r}), \tag{1}$$

where $\mathbf{e_b} = \mathbf{b}/b$ and $\Delta \mathbf{r}$ denotes the radius-vector from the segment center to the observation point, where the stress should be evaluated. Definitions and evaluation methods of the component-specific functions $f_{\alpha\beta}$ will be addressed below.

Under the same assumption $\Delta r \gg \Delta l$, the stress from all dislocation segments of the same dislocation type (*i.e.* with the same Burgers vectors $\mathbf{b}$ and direction vectors $\mathbf{t}$) located inside a small volume $\Delta V$ having a shape of a rectangular prism with the base area $\Delta s$ and the height $\Delta l$ (so that $\Delta V = \Delta s \Delta l$), is equal to

$$\sigma_{\alpha\beta}^{(\Delta V)} = \sum_{i=1}^{N_{\text{disl}}} \sigma_{\alpha\beta}^{(i)} \approx \underbrace{N_{\text{disl}}(\Delta V)}_{n_{\text{disl}}(\mathbf{r}) \cdot \Delta s} \cdot \sigma_{\alpha\beta}^{\text{seg}} = n_{\text{disl}}(\mathbf{r}) \cdot \Delta s \Delta l \cdot b \cdot f_{\alpha\beta} = b \cdot n_{\text{disl}}(\mathbf{r}) f_{\alpha\beta}(\mathbf{e_b}, \mathbf{t}; \Delta \mathbf{r}) \Delta V \tag{2}$$

where $N_{\text{disl}}$ is the total number of dislocation segments inside $\Delta V$. By the derivation of the last expression in (2) we have used the expression (1) for $\sigma_{\alpha\beta}^{\text{seg}}(\Delta l, \mathbf{b}, \mathbf{t})$ and the relation $n_{\text{disl}} = N_{\text{disl}}/\Delta s$ between the total number of dislocation segments $N_{\text{disl}}$ and the dislocation density $n_{\text{disl}}$ (defined as the number of dislocation per unit surface). The relation $n_{\text{disl}} = N_{\text{disl}}/\Delta s$ is valid, strictly speaking, only for dislocations directed along the prism height; however, in the final expression only the total volume of the prism $\Delta V$ is present, what shows that this expression can be applied (under the assumptions outlined above) to dislocations having any direction with respect to the small volume element $\Delta V$.

By integrating over the whole sample volume the stress caused by all dislocations in the system is obtained as the integral

$$\sigma_{\alpha\beta}^{\text{dis}}(\mathbf{r}_0) = b \int_V n_{\text{disl}}(\mathbf{r}) \cdot f_{\alpha\beta}(\mathbf{e_b}, \mathbf{t}; \mathbf{r} - \mathbf{r}_0) \cdot dV \tag{3a}$$

In the general case of $k$ different dislocation types (*i.e.* dislocation with different line directions and Burgers vectors), the summation over these $k$ types should be performed.

$$\sigma_{\alpha\beta}^{\text{dis}}(\mathbf{r}_0) = \sum_{\{k\}} \sigma_{\alpha\beta,k}^{\text{tot}}(\mathbf{r}_0) = \sum_{\{k\}} b_k \cdot \int_V n_{\text{disl}}^{(k)}(\mathbf{r}) \cdot f_{\alpha\beta}^{(k)}(\mathbf{e_b}^{(k)}, \mathbf{t}^{(k)}; \Delta \mathbf{r}) \cdot dV \tag{3b}$$

The basic equations (3a) and (3b) allow the calculation of the total stress field from any known dislocation density if the corresponding interaction functions $f_{\alpha\beta}(\mathbf{e_b}, \mathbf{t}; \Delta \mathbf{r})$ which describe stresses caused by a single dislocation segment are known.

Before we proceed with the evaluation of functions $f_{\alpha\beta}$, we would like to establish some connections between Eqs. (3a) and (3b) and other known formalisms for the evaluation of the dislocation induced stresses, energy of dislocation networks and dislocation dynamics from the dislocation density.

Taking into account that in the linear elasticity theory the functions $f_{\alpha\beta}(\mathbf{e_b}, \mathbf{t}; \Delta \mathbf{r})$ are linear in Cartesian components of vectors $\mathbf{e_b}$ and $\mathbf{t}$, we note that the integral (3a) can be rewritten as a sum of terms proportional to combinations $b_\gamma \cdot t_\delta$. In this form the stress components $\sigma_{\alpha\beta}$ would be expressed as functions of the components of the Nye tensor $\hat{\alpha}$ [Nye1953]. The more general form (3b), which accounts for the presence of many dislocation types, can be related to the so called second order dislocation density tensor $\hat{\alpha}^{II}$ (see, e.g., [Sandfeld2011, Cajic2011]), which is used in continuum dislocation dynamics for the description of evolution of a system of dislocations with different orientations. However, keeping in mind that for experimental applications a dislocation population can be described in many cases by discrete sets of possible Burgers vectors and orientation directions, we shall operate with the forms (3a) and (3b) as more transparent.

We also note, that in this paper we do not consider any dislocation dynamics and will study dislocation induced stresses in polycrystalline materials. For these reasons the products of our dislocation density $n_{\text{disl}}$ with the components of Burgers and direction vectors **b** and **t** should not necessarily obey the restrictions following from the assumption that dislocation lines cannot end inside a crystal and continuity relations of the dislocation dynamics (see corresponding relations for tensors $\hat{\alpha}$ and $\hat{\alpha}^{II}$ in [Nye1953, Sandfeld2010, Sandfeld2011]).

## 2.2 Stress caused by a single dislocation segment

In the simplest case of a straight finite dislocation segment parallel to the *z*-axis of a Cartesian coordinate system, explicit expressions for the stress components are well known [Anderson2017]. Namely, the corresponding stress can be represented as a difference of stresses generated by two semi-infinite dislocations $\hat{\sigma}_{\alpha\beta}^{\text{s-inf}}(\mathbf{r}, z')$ (Fig. 1, left panel).

For a vertical semi-infinite dislocation, there exist several formula sets for the tensor components [Anderson2017, Cai2006], depending on whether the observation point is above, below or in-between the end point of the dislocation segment ($z > z'_B$, $z < z'_A$ or $z'_A < z < z'_B$). We write out the corresponding set for $z < z'_A$, in order to make the paper self-sufficient and for further discussions below:

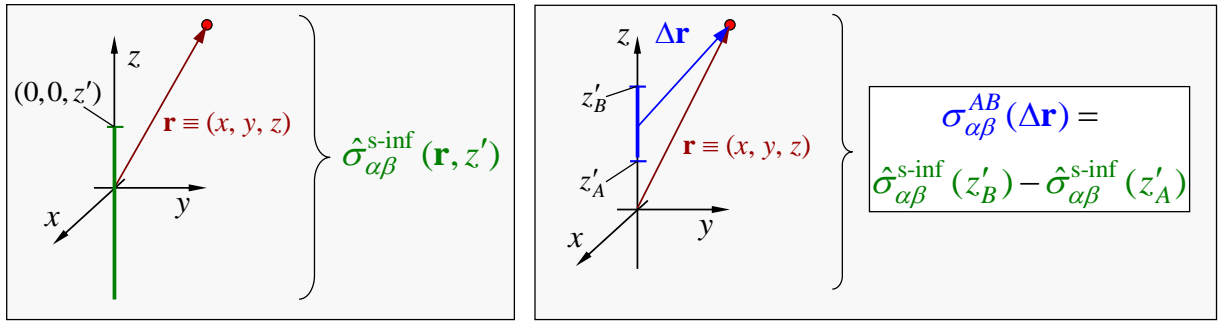

Fig. 1. Semi-infinite dislocation (left) and the finite dislocation segment (right), both having direction vectors **t** = $\mathbf{e}_z$, and the notation for corresponding stress components.

$$\frac{\hat{\sigma}_{xx}}{\sigma_0} = b_x \frac{y}{R_{\lambda+}^2}(1+S_x) + b_y \frac{x}{R_{\lambda+}^2}(1-S_x), \qquad \frac{\hat{\sigma}_{xy}}{\sigma_0} = -b_x \frac{x}{R_{\lambda+}^2}(1-S_y) + b_y \frac{y}{R_{\lambda+}^2}(1-S_x)$$

$$\frac{\hat{\sigma}_{yy}}{\sigma_0} = -b_x \frac{y}{R_{\lambda+}^2}(1-S_y) - b_y \frac{x}{R_{\lambda+}^2}(1+S_y), \qquad \frac{\hat{\sigma}_{xz}}{\sigma_0} = -b_x \frac{xy}{R^3} - b_y \left[\frac{\nu}{R} - \frac{x^2}{R^3}\right] + b_z \frac{y(1-\nu)}{R_{\lambda+}^2} \qquad (4a)$$

$$\frac{\hat{\sigma}_{zz}}{\sigma_0} = (b_x y - b_y x)\left[\frac{2\nu}{R_{\lambda+}^2} + \frac{\lambda}{R^3}\right], \qquad \frac{\hat{\sigma}_{yz}}{\sigma_0} = b_x \left[\frac{\nu}{R} - \frac{y^2}{R^3}\right] + b_y \frac{xy}{R^3} - b_z \frac{x(1-\nu)}{R_{\lambda+}^2}$$

Here the factor $\sigma_0 = \mu/4\pi(1-\nu)$ has the dimensionality of stress and is a combination of the shear modulus $\mu$ and the Poisson ratio $\nu$; functions $R$ and $R_\lambda$ are defined as

$$R^2 = x^2 + y^2 + (z-z')^2, \quad R_{\lambda+}^2 = R(R+\lambda), \qquad (4b)$$

where $\lambda = z' - z$. Functions $S_x$ and $S_y$ are defined via $R$ and $R_\lambda$ by the relations

$$S_x = \frac{x^2}{R^2} + \frac{x^2}{R_{\lambda+}^2}, \quad S_y = \frac{y^2}{R^2} + \frac{y^2}{R_{\lambda+}^2} \qquad (4c)$$

From the presence of the factor $(R + \lambda)$ is the denominator of stress components in (4a), it is clear why the form (4a) is suitable for $z < z'_A$: by definition, $R \geq 0$, so that to avoid a singularity at $R + \lambda = 0$ (i.e., by $R = \lambda$), one should have $\lambda > 0$, implying that $z < z'_A$.

Formulas for the cases $z > z'_B$ and $z'_A < z < z'_B$ can be found in [Anderson2017]; note that application regions for corresponding regularized expressions in [Cai2006] should be interchanged (i.e., *form 1* from [Cai2006] is applicable for $z < z'_A$, and *form 2* - for $z > z'_B$).

For a finite dislocation segment between points A and B (Fig. 1, right panel), stress components should be computed from corresponding expressions for semi-infinite dislocations as

$$\sigma_{\alpha\beta}^{AB}(\Delta\mathbf{r}) = \hat{\sigma}_{\alpha\beta}^{\text{s-inf}}(z'_B) - \hat{\sigma}_{\alpha\beta}^{\text{s-inf}}(z'_A) \qquad (5)$$

To avoid the sometimes encountered incorrect usage of Eqs. (4a) and two other analogous sets, we emphasize the following: All these sets can in principle be used if the singularity in the denominator is not expected, what is the case for $x^2 + y^2 > 0$ (the observation point is not located at the $z$-axis). However, one should keep in mind that *absolute values* of the stress of a *semi-infinite* dislocations, provided by these three sets of expressions will differ by some constants, because these sets are *integral* solutions for a *semi-infinite* dislocation. Hence Eqs. (4a) and analogous formulas for other $z$-regions are suitable *only* for the calculation of *stress differences* from these semi-infinite segments, but *not* of their absolute values. In our case this is not a limitation, because we calculate stresses of *finite* segments employing Eq. (5). Hence we have a free choice between different sets, which allows us to avoid artificial singularities in corresponding $z$-regions, what is especially convenient in numerical simulations.

For a dislocation segment with arbitrary orientations of **t** and **b**, several different formulas - in coordinate-bonded and coordinate-independent forms - are available (see [Anderson2017, Devincre1992, Arsenlis2007]). We have chosen the coordinate-independent expression

$$\sigma_{\alpha\beta}^{AB} = \frac{\mu}{\pi R_t^2} \left( [\mathbf{bR}_t\mathbf{t}]_{\alpha\beta}^{\text{sym}} - \frac{1}{(1-\nu)}[\mathbf{btR}_t]_{\alpha\beta}^{\text{sym}} - \frac{1}{2(1-\nu)}(\mathbf{b},\mathbf{R}_t,\mathbf{t})(\delta_{\alpha\beta} + t_\alpha t_\beta + \Phi_{\alpha\beta}) \right) \Big|_A^B \qquad (6)$$

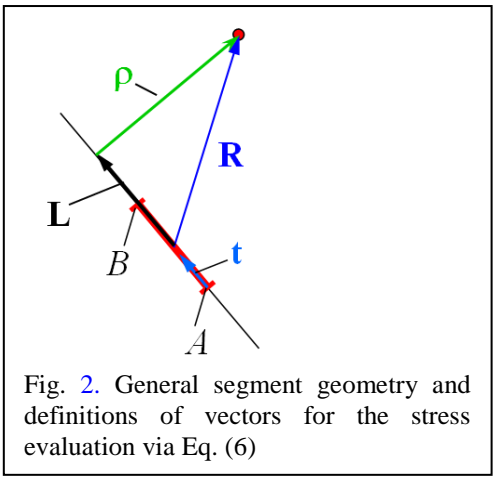

Fig. 2. General segment geometry and definitions of vectors for the stress evaluation via Eq. (6)

from [Devincre1992], which can be readily used in numerical simulations. In this expression, vector $\mathbf{R}_t$ is defined via the segment direction vector **t** and the vector **R** between the middle point of the segment and the observation point as $\mathbf{R}_t = \mathbf{R} + R \cdot \mathbf{t}$ (see Fig. 2). Functions $\Phi_{\alpha\beta}$ are expressed in terms of vectors **L** (with $L = \mathbf{R} \cdot \mathbf{t}$) and $\boldsymbol{\rho} = \mathbf{R} - L \cdot \mathbf{t}$ defined in Fig. 2 as

$$\Phi_{\alpha\beta} = \frac{2}{R_t^2}\left( \rho_\alpha R_{t,\beta} + \rho_\beta R_{t,\alpha} + \frac{L}{R} R_{t,\alpha} R_{t,\beta} \right) \quad (7a)$$

Three-vectors operators in Eq. (7a) are defined as

$$[\mathbf{abc}]_{\alpha\beta}^{\text{sym}} = \frac{1}{2}\left([\mathbf{a}\times\mathbf{b}]_\alpha c_\beta + [\mathbf{a}\times\mathbf{b}]_\beta c_\alpha\right), \quad (\mathbf{a},\mathbf{b},\mathbf{c}) = \begin{vmatrix} a_x & a_y & a_z \\ b_x & b_y & b_z \\ c_x & c_y & c_z \end{vmatrix} \quad (7b)$$

Expressions for interaction functions $f_{\alpha\beta}(\mathbf{e_b},\mathbf{t};\Delta\mathbf{r})$ can be obtained from the formulas for the stress components of a dislocation segment in two different ways.

In the first case - when derivatives of stress components for a semi-infinite segment $\hat{\sigma}_{\alpha\beta}^{\text{s-inf}}$ can be evaluated analytically - we can use the relation $z'_B = z'_A + \Delta l$ and rewrite Eq. (5) as

$$\sigma_{\alpha\beta}^{AB} = \hat{\sigma}_{\alpha\beta}^{\text{s-inf}}(z'_B) - \hat{\sigma}_{\alpha\beta}^{\text{s-inf}}(z'_A) = \hat{\sigma}_{\alpha\beta}^{\text{s-inf}}(z'_A + \Delta l) - \hat{\sigma}_{\alpha\beta}^{\text{s-inf}}(z'_A) \approx \left.\frac{d\hat{\sigma}_{\alpha\beta}}{dz'}\right|_{z'_A} \Delta l = \frac{1}{b}\left.\frac{d\hat{\sigma}_{\alpha\beta}}{dz'}\right|_{z'_A} \cdot b\Delta l \quad (8).$$

Comparison with Eq. (1) leads to the expression

$$f_{\alpha\beta}(\mathbf{e_b},\mathbf{t}) = \frac{1}{b}\left.\frac{d\hat{\sigma}_{\alpha\beta}}{dz'}\right|_{z'_A} \quad (9a)$$

However, in almost all application-relevant cases, analytical derivatives of the segment stress are not available, or are so complicated that their usage destroys all common advantages of analytical formulas. In this situation, we use an obvious transformation, resulting in the following expression for $f_{\alpha\beta}(\mathbf{e_b},\mathbf{t};\Delta\mathbf{r})$:

$$\sigma_{\alpha\beta}^{AB} = \frac{\sigma_{\alpha\beta}^{AB}}{\Delta l \cdot b} \cdot \Delta l \cdot b \quad \Rightarrow \quad f_{\alpha\beta}(\mathbf{e_b},\mathbf{t}) = \frac{\sigma_{\alpha\beta}^{AB}}{\Delta l \cdot b} \quad (9b)$$

Here we would like to comment on the relation between the evaluation of the dislocation-induced stress in the widely used discrete dislocation dynamics (see references in the Introduction) and our method. DDD handles systems which can be described as a collection of *single* (well separated) dislocations Our formalism is applicable to systems which are described using the concept of the dislocation *density*, so that a direct comparison with DDD is not possible. However, as we use the same initial equations (Eqs. (4) - (7)) as in numerous papers devoted to DDD and cited above, our method has the same correctness as the DDD, namely, it is valid in frames of the mesoscopic elasticity theory of dislocations.

### 2.3. Numerical evaluation of the dislocation induced stress

The integrals (3a) or (3b) cannot be evaluated analytically in all practically relevant cases already due to the very complicated expressions for the interaction functions $f_{\alpha\beta}(\mathbf{e_b},\mathbf{t};\Delta\mathbf{r})$, not to mention the important circumstance that the dislocation density can rarely be approximated by an analytical function. For this reason, we have to resort to numerical methods to evaluate these integrals, approximating them as sums over all finite elements used for the discretization of our system. For the dislocations of the same type, the integral (3a) will thus be converted into the sum

$$\sigma_{\alpha\beta}^{\text{dis}}(\mathbf{r}_0) = b\sum_i n_{\text{disl}}(\mathbf{r}_i) \cdot f_{\alpha\beta}(\mathbf{e_b},\mathbf{t};\mathbf{r}_0 - \mathbf{r}_i) \cdot \Delta V_i \quad (10)$$

If several dislocation types are present, we have to perform the summation (10) for each dislocation type separately using corresponding densities $n$ and interaction functions $f$ and then sum up all these partial stresses.

Three important remarks are in order. First, the evaluation of (10) via the direct summation over all volume elements $\mathbf{r}_i$ for all observation points $\mathbf{r}_0$ (also located in all finite elements) leads to an operation count $\sim N^2$, where $N$ is the total number of discretization elements (cells). Hence this method is too slow for any realistic 3D model, which usually involves about $10^5$ or more discretization cells. Fortunately, the interaction functions $f_{\alpha\beta}(\mathbf{e_b}, \mathbf{t}; \Delta\mathbf{r})$ are translationally invariant, *i.e.* depend only on the difference $\Delta\mathbf{r} = \mathbf{r}_0 - \mathbf{r}_i$ between the radius-vectors of the observation point $\mathbf{r}_0$ and the volume element $\mathbf{r}_i$. This makes the expression (10) to a numerical convolution, which can be easily evaluated via the fast Fourier transformation (FFT); see, e.g., Chap. 13 in [Press1992] for details. Evaluation via FFT requires only $\sim N \cdot \log N$ operations, enabling to handle systems with up to $N \sim 10^7$ - $10^8$ discretization cells on a standard CPU. However, one should keep in mind that a prerequisite for the FFT usage is a translationally invariant (regular) discretization grid.

Second, the accuracy of the stress evaluation via the finite element summation (10) should be discussed. The main question here is how accurate is the approximation of the stress $\sigma_{\alpha\beta}(\mathbf{r}_i \to \mathbf{r}_0) = b \cdot n_{\text{disl}}(\mathbf{r}_i) \cdot f_{\alpha\beta}(\mathbf{e_b}, \mathbf{t}; \mathbf{r}_0 - \mathbf{r}_i) \cdot \Delta V_i$ generated on the finite element with the center at $\mathbf{r} = \mathbf{r}_0$ by dislocations within the element located at $\mathbf{r} = \mathbf{r}_i$ for *small* distances $\Delta\mathbf{r} = \mathbf{r}_0 - \mathbf{r}_i$ (nearest neighboring cells).

The expression (1) by itself can not cause any numerical errors, because if we evaluate interaction functions $f_{\alpha\beta}$ using the method (9b), then Eq. (1) is exact for *all* distances $\Delta r$, including those which are smaller than the segment length $\Delta l$. Hence the only source of numerical errors is the usage of the same distance $\Delta\mathbf{r} = \mathbf{r}_0 - \mathbf{r}_i$ for all points within the elements $\Delta V_0$ and $\Delta V_i$ by computing the stress $\sigma_{\alpha\beta}(\mathbf{r}_i \to \mathbf{r}_0)$. This is the standard problem concerning the accuracy of the multipole expansion of any interaction potential. In this formalism, retaining only the term $b \cdot n_{\text{disl}}(\mathbf{r}_i) \cdot f_{\alpha\beta}(\mathbf{e_b}, \mathbf{t}; \mathbf{r}_0 - \mathbf{r}_i) \cdot \Delta V_i$ by the evaluation of the integral (3a) via (10) is equivalent to usage of the lowest non-zero term of the multipole expansion (because this term is independent on the finite element shape). Contribution of higher multipole terms can be minimized by choosing finite elements (discretization cells) whose shape is as isotropic as possible. In our case of a regular rectangular grid this means that cubic cells should be preferred in simulations.

However, in many situations the usage of cells even with approximately equal sizes in all directions would lead to prohibitively long computation times. Typical examples are polycrystalline layers with grains whose typical lateral size is much larger or much smaller than the layer thickness, systems with dislocation densities varying rapidly in one direction (often the growth direction) and nearly constant in other directions etc.

For this reason we have developed a method which allows a sufficiently accurate evaluation of interaction coefficients $f_{\alpha\beta}(\mathbf{e_b}, \mathbf{r}_0 - \mathbf{r}_i)$ between strongly shape-anisotropic cells without using higher multipole moments. To achieve this goal, we divide a target cell and source cells within several nearest-neighbors shells into subcells; the shape of these subcells should be as close to cubical as possible (see Fig. 3). Then we compute the interaction coefficients between the corresponding source and the target cells as the sum of interactions between subcells contained in these cells. For remote cells, no subdivision is necessary. This procedure (evaluation of the interaction coefficients) should be performed only once for the given simulation geometry so that the initial loss in computational time is more than compensated by the increased accuracy and smaller number of discretization cells. This method is used in all systems simulated below, especially by modeling of polycrystalline layers in Sec. 6.

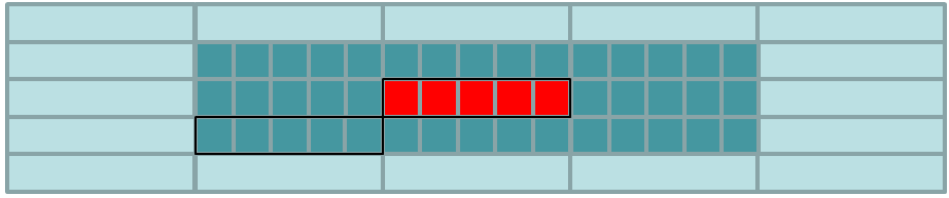

Fig. 3. To the calculation of interaction coefficients for non-cubic cells: the target cells (red) and several shells of the source cells surrounding the target cell are subdivided into nearly cubic cells.

The last problem is the singular behaviour of the expressions (4) and (6) if the distance between the source and target cells tends to zero: $\Delta r \to 0$. In general, this is a non-trivial problem, which requires the introduction of the regularized (*i.e.* non-singular in the limit $\Delta r \to 0$) expressions for the dislocation stress when the contribution from *each* dislocation has to be taken into account separately. Detailed discussion of corresponding expressions is given in [Cai2006]. In our case this singularity requires a separate treatment of self-interaction terms, *i.e.* the contribution to the stress in each cell coming from dislocations within the same cell. This contribution can be estimated as integrals from the expressions (4) over the small volume $\Delta V$ of a discretization cell. It is easy to see that for small distances $R$ the stress components (4) diverge as $1/R$, so that corresponding integrals over the volume including the point $R = 0$ converge. From the point of view of numerical simulations this means that self-interaction terms tend to zero for $\Delta V \to 0$ (*i.e.* when the discretization cell size decreases). In addition, contributions from most tensor components vanish by such an integration, because $\sigma_{xx}$, $\sigma_{yy}$, $\sigma_{zz}$, and $\sigma_{xy}$, are odd functions of $x$ and $y$.

In the results presented below, we have always verified (by halving the cell size) that further discretization refinement did not lead to noticeable changes in the result.

## 3. Stress induced by a continuous distribution of dislocations of the same type

Before considering a realistic example of a dislocation distribution in a polycrystalline film, we would like to present results for a dislocation induced stress in a system, where all dislocations have the same type. Although physically quite unlikely, this example is highly instructive, allowing us to analyze corresponding stresses in a relatively simple case and to demonstrate the efficiency of our computational method.

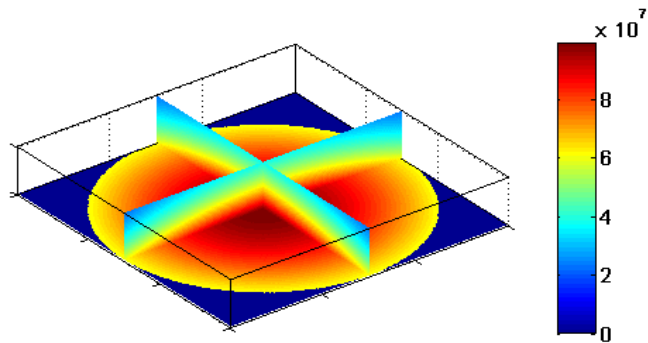

Fig. 4. 3D image of the sample dislocation density (11) using throughout this section.

As a sample system we have chosen a disk with the diameter $D = 2.5$ cm and thickness $h = 5$ mm made of a material with parameters typical for semiconductor crystals: Young modulus $E = 200$ GPa, Poisson number $\nu = 0.2$ (leading to the shear modulus $\mu \approx 83$ GPa) and a hexagonal lattice with lattice constants $a = c = 5$ Å. Further, we have assumed that the dislocation density is site-dependent, decaying from the disk bottom to its top and from the center to the side borders as

$$\rho(z,r) = A_{\max} \cdot \exp\left\{-\frac{r^2}{2l_r^2}\right\} \cdot \exp\left\{-\frac{z}{l_z}\right\} \qquad (11)$$

where the $z$-axis is perpendicular to the disk plane. The maximal value of this density was set to $A_{\max} = 10^8$ cm$^{-2}$, the radial and the vertical decay lengths - to $l_r = h$ and $l_z = D/2$. 3D picture of the density (11) with these parameters is shown in Fig. 4. The disk volume was discretized into $N_x \times N_y \times N_z = 250 \times 250 \times 50 \approx 3.1 \times 10^6$ cells.

In the first example (Fig. 5), we show spatial patterns of the stress components (computed using (10)) induced by vertical edge dislocations with the spatial density (11) and identical Burgers vectors $\mathbf{b} = a \cdot \mathbf{e}_x$. As for all dislocations the only the $b_x$-component of the Burgers vector is not zero, spatial symmetry patterns in Fig. 5 can be easily explained from the symmetry of the terms containing $b_x$ in (4). In particular, the only component which does not change sign in the disk plane, is the $\sigma_{yz}$-component, because the contributions to $\sigma_{yz}$ from points across the disk (lying in the same $xy$-plane) are accumulated, according to (4). For all other components, these contributions are odd functions of $x$ or $y$, so that the total stress components change sign within the disk plane. Another interesting feature clearly visible in Fig. 5 is the relatively weak dependence of the diagonal and $\sigma_{xy}$ components on the distance from the disk bottom ($z$-coordinate), although the dislocation density (11) is strongly $z$-dependent. This behaviour is the direct consequence of the long-range nature of the elastic stresses (see Eqs. (4)).

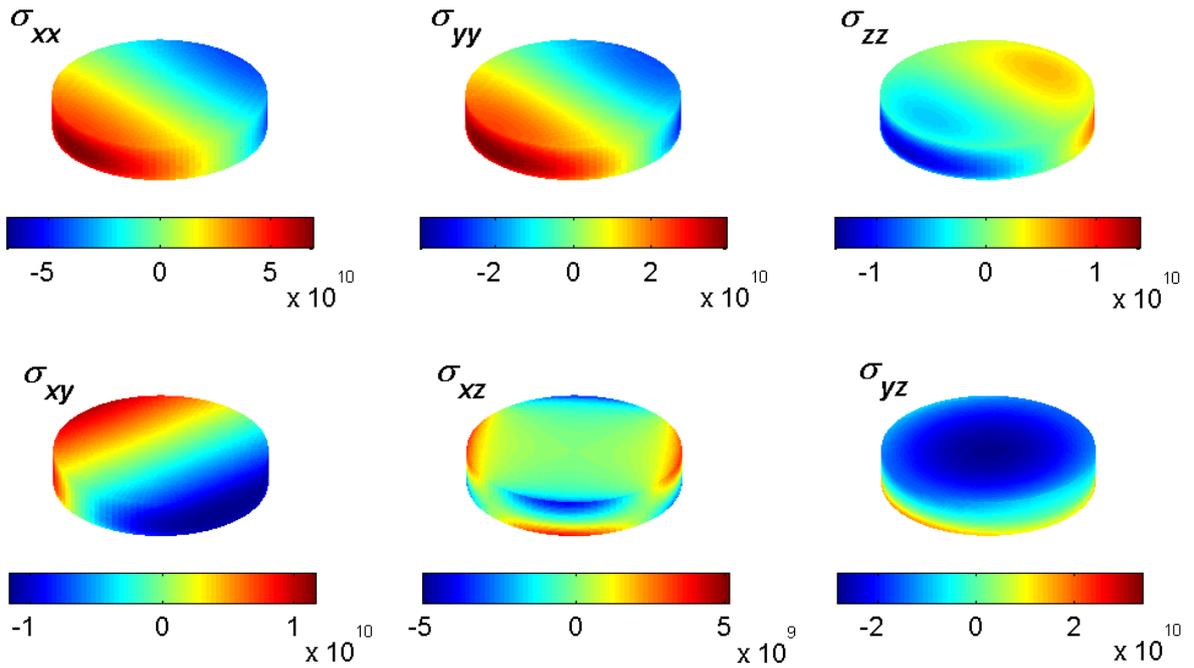

Fig. 5. Stress from edge dislocations with the spatial density (11), all having the same Burgers vectors $\mathbf{b} = a \cdot \mathbf{e}_x$. Symmetry patterns following from the expressions (4) can be clearly recognized.

In the second example we consider the stress induced by dislocations with the same density (11), but with the Burgers vectors $\mathbf{b} = \mathbf{a} + \mathbf{c}$: $b_x = a/2$, $b_y = \sqrt{3}a/2$, $b_z = c$ and the direction vectors $\mathbf{t}$: $(t_x = 0, t_y = 1/2, t_z = \sqrt{3}/2)$ - the so called mixed dislocations. Stress components for this case are shown in Fig. 6. Due to the presence of all components of the Burgers vector and two non-zero components of the direction vector, the stress pattern is qualitatively different from that from the previous example. In particular, this pattern is now 'tilted', and neither of the stress components possesses any simple spatial symmetry, in contrast to the case of the vertical edge dislocations.

Maximal stress values in these examples are extremely high, reaching ~ 50 GPa. Such high values are obviously due to the assumption that all dislocations in the disk have the same Burgers and direction vectors. The problem how the typical stress value change when we assume a polycrystalline sample structure with different dislocations in each crystallites, will be addressed in detail in the next sections.

Finally we emphasize that stress components shown in Fig. 5 and 6 are computed numerically using (10) with the interaction functions derived from expression (4), (5) and (6). This means that elastic boundary conditions at the free surfaces of a finite body are *not* taken into account. In this sense, stress components presented in the two figures above correspond to the physical situation where the disk containing dislocations with the spatial density (11) is ‚embedded' into an infinite elastic continuum with the same elastic constants.

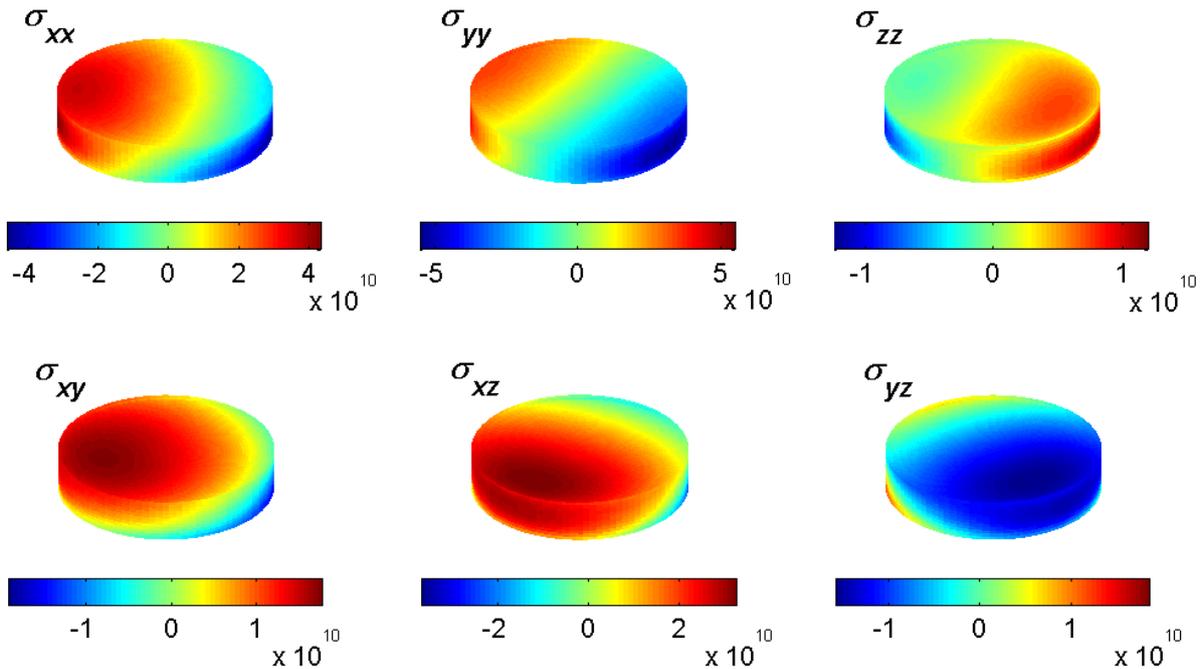

Fig. 6. Stress from mixed dislocations with the spatial density (11), the same Burgers vectors $\mathbf{b} = \mathbf{a} + \mathbf{c}$ ($b_x = a/2$, $b_y = \sqrt{3}a/2$, $b_z = c$) and direction vectors $\mathbf{t}$: ($t_x = 0$, $t_y = 1/2$, $t_z = \sqrt{3}/2$).

For a real systems like a disk with free surfaces one should solve the corresponding problem of the elasticity theory with dislocation-induced stresses considered as external loads and with proper boundary conditions. In several simple cases, corresponding solution for a single dislocation can be obtained using the image method (see, e.g. [Anderson2017]). In a more general case, sophisticated numerical methods should be applied, in order to solve this problem so efficiently that the solution can be updated at each time step of the dislocation dynamics - see corresponding discussion in, e.g., [ElAwady2008] and [Weinberger2009]. If we are interested only in the stress pattern for a given dislocation distribution, it is possible to input the stress components computed from (10) as external stresses into a finite-element software for mechanical simulations (like ANSYS Mechanical, COMSOL etc.), which will handle this problem with an embedded solver applying proper boundary conditions. We shall not discuss this issue for the two simple examples above, leaving the more detailed analysis till Sec. 5 and 6, where more realistic cases will be considered.

## 4. Implementation of periodic boundary conditions for simulations of thin films

Polycrystalline films employed in technological applications are normally too large to enable direct simulation of the dislocation-induced stress in an entire film. As an example, we can estimate the number of finite elements required for the discretization of a practically interesting system as follows. The diameter of wafers $D_w$ used in the semiconductor industry is about several inches. Taking for our estimate $D_w = 10$ cm and considering a film with the average crystallite size of $d_{cr} = 0.1$ mm, we find that a semiconductor film grown on such a wafer contains $N_{cr} \sim (D_w/d_{cr})^2 = 10^6$ crystallites. Each crystallite should be discretized sufficiently fine in the disc plane (*xy*-plane), meaning that the number of cells in each direction should be not less than $N_x \approx N_y \sim 10$. This leads to the number of discretization cells in only one in-plane discretization layer $N_{\text{in-plane}} \sim 10^8$ cells, and this number still needs to be multiplied by the number of discretization layers in the direction perpendicular to film plane. Thus, simulation of an entire system with an adequate spatial resolution is unrealistic.

Taking into account that only a relatively small part of a real system can be simulated, we need to implement periodic boundary conditions (PBC) in the film plane - otherwise the influence of artificial boundaries would be too strong. We note that for a real system the layer edges are far away from the simulated volume which has been cut out of this system.

The concept of PBC means that the system to be simulated is periodically continued in all directions where these conditions are applied (in our case in both lateral directions). The stress at any target point $\mathbf{r}_0$ from dislocations around the source point $\mathbf{r}_i$ should thus be calculated by the summation of contributions from all dislocation segments contained in the small volume $\Delta V_i$ (with the center at $\mathbf{r}_i$) and from *all* replicas of this volume appearing due to its periodic continuation in each lateral direction according to PBC (Fig. 7).

Thus the stress induced at the point $\mathbf{r}_0$ by dislocations within the volume element $\Delta V(\mathbf{r}_i) \equiv \Delta V_i$ should be evaluated as the sum (2), extended over all periodic system replica shown in Fig. 7 (left panel):

$$\sigma_{\alpha\beta}^{(\text{PBC})}\left(\mathbf{r}_i \to \mathbf{r}_0\right) = \sum_{k,l=-\infty}^{\infty} \sum_{m=1}^{N_{\text{disl}}(\Delta V_i)} \sigma_{\alpha\beta}^{\text{seg}}\left(\mathbf{r}_{i,m}^{(kl)} - \mathbf{r}_0\right) = \sum_{k,l=-\infty}^{\infty} \sum_{m=1}^{N_{\text{disl}}} b \cdot \Delta l \cdot f_{\alpha\beta}\left(\mathbf{r}_{i,m}^{(kl)} - \mathbf{r}_0\right) \approx$$

$$\approx b \cdot \Delta l \sum_{k,l=-\infty}^{\infty} f_{\alpha\beta}\left(\mathbf{r}_i^{(kl)} - \mathbf{r}_0\right) = b \cdot n_{\text{disl}} \cdot \Delta V_i \cdot F_{\alpha\beta}^{\text{PBC}}\left(\mathbf{r}_i - \mathbf{r}_0\right) \qquad (12)$$

Each interaction function $F_{\alpha\beta}^{\text{PBC}}$ in (12) is an infinite sum of 'initial' interaction functions $f_{\alpha\beta}$

$$F_{\alpha\beta}^{\text{PBC}}\left(\mathbf{r}_i - \mathbf{r}_0\right) = \sum_{k,l=-\infty}^{\infty} f_{\alpha\beta}\left(\mathbf{r}_i^{(kl)} - \mathbf{r}_0\right) \qquad (13)$$

The most effective method to perform the summation (13) is based on the Fourier transform (FT) of the interaction coefficients $f_{\alpha\beta}(\mathbf{r})$. This method can be employed, because the simulated system is now periodic in-plane. A well known problem arising by the FT usage in this situation is due a limited number of Fourier components available in numerical modeling. Hence the Fourier spectrum of the interaction functions is cut off at some finite wave vector $\mathbf{k}$, what leads to unphysical oscillations of the interaction in the real space. To solve this problem, the so-called Ewald methods [Hockney1988] are used, which employ a decomposition of the interaction in a short-range and a long-range part. These parts are constructed in the way which ensures a rapid decay of Fourier components of the long-range part in the **k**-space, so that the spectrum cut-off at some finite but large $k$ does not significantly affect the result. This class of methods is very widely used in Coulomb, gravitational [Hockney1988] and dipolar [DeBell2000] systems. However, the interaction functions between dislocations have a very complicated analytical form. For this reason the development and programming effort for the construction of the corresponding Ewald decomposition and evaluation of long-range Fourier harmonics is very high already for a 2D system of straight dislocations [Barts1995], not to mention a general 3D case considered here.

Implementation of a fast multipole method (FMM), which asymptotic operation count ~ $N$ is even better than ~$N \cdot \log N$ for the Ewald method (however, the prefactor in FMM is much larger), is also very complicated [Wang1995, Arsenlis2007].

For these reasons we have implemented the third (and most straightforward) possibility to evaluate the sums over replica (13) with any prescribed accuracy by performing the summation over the expanding replica shells (see Fig. 7, left panel) and cutting the sum at some particular shell $n_{sh}$.

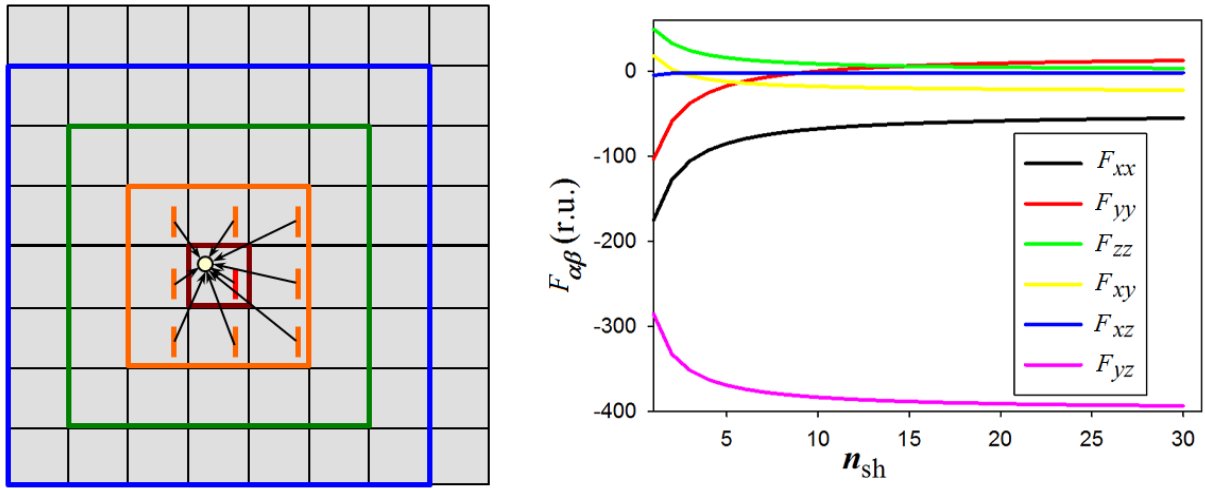

Fig. 7. Construction of the sum (13) for the evaluation of the interaction functions $F_{\alpha\beta}^{PBC}$ for a system with periodic boundary conditions using the summation over subsequent expanding shells. **Left** panel: initial system (small dark red square) with a target point (open circle) and a source dislocation which contribution should be evaluated (red); next shells are shown with color expanding squares. Images of the source dislocation (from initial system) in the replica belonging to the first shell are shown in orange. **Right** panel: Convergence of the interaction functions $F_{\alpha\beta}^{PBC}$ with increasing number of shells.

The convergence of the sums (13), when the summation is performed over replica shells requires a separate discussion. For segments of a straight dislocation with $\mathbf{t} = \mathbf{e}_z$, where expressions (4) can be used, the expansion of the stress components $\sigma_{\alpha\beta}^{seg}(\Delta l, r)$ in a small parameter $\Delta l / r \ll 1$ leads to different asymptotic behaviour for different tensor components. For $\sigma_{xx}$, $\sigma_{yy}$, $\sigma_{zz}$, and $\sigma_{xy}$, we obtain the relatively slow decay with the distance from the segment $r$, namely $\sigma_{\alpha\beta}^{seg}(\Delta l, r) \sim \Delta l / r^2$, so that in a 2D system corresponding sums could diverge logarithmically. However, these stress components change their sign in the $xy$-plane, being *odd* functions of $x$ or $y$ (or both of them), so that sums over shells for these components converge. Components $\sigma_{xz}$ and $\sigma_{yz}$ are *even* functions of $x$ and $y$, but the analysis shows that they have the asymptotic behaviour $\sigma_{\alpha\beta}^{seg}(\Delta l, r) \sim \Delta l / r^3$, again leading to the convergence of the sums over shells in 2D. In a general case of an arbitrary dislocation segment described, e.g., by Eq. (6), each stress component in the global coordinate system is a linear combination of components written in a local (segment-attached) coordinate system, so that corresponding sums over replica shells should also converge.

Numerical results of the convergence test are shown in Fig. 7 (right panel), where the PBC interaction coefficients $F_{\alpha\beta}^{PBC}$ for a vertical dislocation segment are presented as functions of the number of shells $n_{sh}$ used in the summation (13). It can be seen that the sums over replica shells clearly converge, and that 15 to 20 shells provide the accuracy better than 3% for all stress components.

We also point out that interaction functions $F_{\alpha\beta}^{PBC}$ should be evaluated only once at the beginning of simulations. For these reason this method is very efficient, if many dislocation configurations for one and the same system should be studied. In particular, the method is very well suited for simulations of the dislocation dynamics in the dislocation density formalism.

## 5. Relaxation of stresses taking into account elastic boundary conditions

In this Section we return to the question of the determination of the total stress in a system containing dislocations under consideration of correct elastic boundary conditions. 'Internal' stress components $\sigma_{\alpha\beta}$ in this case obey the standard equations of the linear elasticity theory which describes the equilibrium of each small volume segment under the action of internal stresses and external volume forces with the given force density (force per unit volume) $\mathbf{f}^{ext}$ (see e.g. [Landau1970, Sadd2009] or any textbook on the elasticity theory). From these equations and the Hook's law for the relation between the stress $\sigma_{\alpha\beta}$ and deformation tensor $u_{\alpha\beta} = (\partial u_\alpha / \partial \beta + \partial u_\beta / \partial \alpha)/2$ (here $\alpha, \beta = x, y, z$ and $u_\alpha$ denote the components of the site-dependent deformation vector $\mathbf{u}$), the basic equation describing the equilibrium state of an elastic body under the action of external volume forces can be derived [Landau1970, § 7]:

$$A \text{ grad div } \mathbf{u}(\mathbf{r}) - B \text{ rot rot } \mathbf{u}(\mathbf{r}) + \mathbf{f}^{ext}(\mathbf{r}) = 0 \quad (14)$$

where the coefficients $A = E(1-\nu)/(1+\nu)(1-2\nu)$ and $B = E/2(1+\nu)$ are expressed via the Young modulus $E$ and the Poisson ratio $\nu$.

'External' forces in (14) are caused in our case by the dislocation-induced stress computed via (3a) and (3b):

$$f_\alpha^{ext} = \sum_{\beta=x,y,z} \frac{\partial \sigma_{\alpha\beta}^{dis}}{\partial \beta} = \frac{\partial \sigma_{\alpha x}^{dis}}{\partial x} + \frac{\partial \sigma_{\alpha y}^{dis}}{\partial y} + \frac{\partial \sigma_{\alpha z}^{dis}}{\partial z} \quad (\alpha = x, y, z), \quad (15)$$

Next we need to established boundary conditions (BCs) for the solution $\mathbf{u}(\mathbf{r})$ of Eq. (14). Assuming a coordinate system with $x$- and $y$-axes in the film plane, we obtain the displacement BCs for the vertical (i.e. perpendicular to the film plane) boundaries of the simulation area in the form

$$u(x=0, y, z) = u(x=L_x, y, z) \text{ and } u(x, y=0, z) = u(x, y=L_y, z) \quad (16a)$$

Top and bottom surfaces of the film are in our case *free* surfaces, parallel to the $xy$-plane and thus perpendicular to the $z$-axis. For such a surface, the following BCs of the traction type should be applied: all '$z$-contained' components of the total stress ($\sigma_{xz}$, $\sigma_{yz}$, $\sigma_{zz}$) must be zero at $z = 0$ and $z = h$, where $h$ is the film thickness. In our case, i.e. in presence of the external load expressed in form of the dislocation-induced stress $\hat{\sigma}^{dis}$, we obtain:

$$\begin{aligned}\sigma_{\alpha z}(\mathbf{r})\big|_{z=0} + \sigma_{\alpha z}^{dis}(\mathbf{r})\big|_{z=0} = 0, \quad \alpha = x, y, z \\ \sigma_{\alpha z}(\mathbf{r})\big|_{z=h} + \sigma_{\alpha z}^{dis}(\mathbf{r})\big|_{z=h} = 0, \quad \alpha = x, y, z\end{aligned} \quad (16b)$$

Using the relation between stress and deformation tensors (Hooke's law), and the definition of the deformation tensor $u_{\alpha\beta}$ via derivatives of displacement components $\partial u_\alpha / \partial \beta$ (see above) these traction-type BCs lead to the following displacement-type BCs:

$$\frac{1}{2}\left(\frac{\partial u_x}{\partial z}+\frac{\partial u_z}{\partial x}\right)\bigg|_{z=0} = -\frac{1+\nu}{E}\sigma_{xz}^{\text{dis}}\bigg|_{z=0} \equiv \varphi_{xz}^{(0)}, \quad \frac{1}{2}\left(\frac{\partial u_y}{\partial z}+\frac{\partial u_z}{\partial y}\right)\bigg|_{z=0} = -\frac{1+\nu}{E}\sigma_{yz}^{\text{dis}}\bigg|_{z=0} \equiv \varphi_{yz}^{(0)},$$

$$(1-\nu)\frac{\partial u_z}{\partial z}+\nu\left(\frac{\partial u_x}{\partial x}+\frac{\partial u_y}{\partial y}\right)\bigg|_{z=0} = -\frac{(1+\nu)(1-2\nu)}{E}\sigma_{zz}^{\text{dis}}\bigg|_{z=0} \equiv \varphi_{zz}^{(0)}$$

(16c)

and the same for $z = h$; dimensionless functions $\varphi_{\alpha z}^{(0)}$ (and $\varphi_{\alpha z}^{(h)}$ for $z = h$) are introduced in (16c) to shorten the notation.

After the discretization of the simulated area into $N = N_x \times N_y \times N_z$ cells the system of 3 partial differential equations (14) is converted into a system of $3N$ linear equations for $3N$ variables $u_{ijk}^{(\alpha)} \equiv u^{(\alpha)}(x_i, y_j, z_k)$, $\alpha = x, y, z$. For the numerical solution of this system we use the so called relaxation method [Press1992]: instead of solving the original equation (14), we solve the discretized 'evolution' equation

$$\frac{\partial \mathbf{u}(\mathbf{r},t)}{\partial t} = A\ \text{grad div}\ \mathbf{u}(\mathbf{r},t) - B\ \text{rot rot}\ \mathbf{u}(\mathbf{r},t) + \mathbf{f}^{\text{ext}}(\mathbf{r}) \qquad (17)$$

using the explicit evolution algorithm, which updates the **u**-values with 'time' according to the scheme $\mathbf{u}_{ijk}(t+\Delta t) \equiv \mathbf{u}_{ijk}(t) + \Delta t \cdot \mathbf{v}_{ijk}(\mathbf{u}_{i\pm 1,j\pm 1,k\pm 1}, \mathbf{f}_{ijk})$. We emphasize, that the presence of the time derivative on the left-hand side of (17) does not mean that we are studying the system dynamics; solution of Eq. (17) is merely a method to find the equilibrium state described by Eq. (14). Due to the discrete representation of partial derivatives in Eqs. (14) or (17), the functions $\mathbf{v}_{ijk}$ depend on the external stress at the same spatial point ($ijk$) and on the displacement values **u** at the previous time step and at neighboring spatial points.

From the theory of partial differential equations it is known that the 'initial' distribution $\{\mathbf{u}_0(\mathbf{r})\} = \{\mathbf{u}(\mathbf{r},\ t = 0)\}$ of this evolution equation converges to the equilibrium state $\mathbf{u}(\mathbf{r}, t \to \infty) \to \mathbf{u}_{\text{eq}}(\mathbf{r})$, which is the solution of the original equation (14), if the time step is sufficiently small to ensure the stability of the solution scheme; for more details see, e.g., [Press1992], Chap. 19. The semi-quantitative von-Neumann analysis leads to the estimation of the maximal allowed time step as $\Delta t \approx \min(\Delta x, \Delta y, \Delta z)^2 / (5A - 3B)$; this estimation has been proven to be nearly optimal and was used in all our simulations.

By the solution of this system, BCs (16a) and (16c) should be taken into account. Inclusion of BCs (16a) on side surface is straightforward. In order to take into account BCs (16c), we have to discretize them using additional layers below (i.e. for $k = 0$) and above (for $k = N_z + 1$) the simulated film. For example, the first BC from (16c) can then be rewritten as

$$\frac{1}{2}\left(\frac{u_{i,j,2}^{(x)} - u_{i,j,0}^{(x)}}{2\Delta z} + \frac{u_{i+1,j,1}^{(z)} - u_{i-1,j,1}^{(z)}}{2\Delta x}\right) = \varphi_{xz}^{(0)}(x_i, y_j) \qquad (18a)$$

This relation is used to update the values of $u_x$ at the bottom layer with $k = 0$ after each time step according to

$$u_{i,j,0}^{(x)} = u_{i,j,2}^{(x)} + 2\Delta z\left(\frac{u_{i+1,j,1}^{(z)} - u_{i-1,j,1}^{(z)}}{2\Delta x} - 2\varphi_{xz}^{(0)}(x_i, y_j)\right) \qquad (18b)$$

In the same manner, the second and third BCs from (16c) are used to update the sets $u_y$ and $u_z$ at the bottom layer ($k = 0$) correspondingly. Displacement values for $k = N_z + 1$ are updated using relations analogous to (18b), but derived from BCs for $z = h$.

# 6. Results and discussion

With the methods described in previous sections, we have studied dislocation-induced stress in polycrystalline films of the semiconductor GaN. The importance of this material and the relevance of dislocation-induced stress in films consisting of GaN was discussed in details in the Introduction. Here we have performed systematic investigations of this stress, studying its dependence especially on the average grain size within the corresponding film.

To obtain the results presented below, we have simulated a region of a GaN layer with the thickness $h = 0.5$ mm (typical for applications of this material), having the square in-plane shape. The lateral size of the simulated area $L$ was varied between 0.01 mm and 10 mm. To study systems with various grain sizes, we have generated the same average number of crystallites $N_{cr} = 100$ within the simulated area, so that the average grain size was varied in the region $D_{av} = 0.001 - 1.0$ mm.

Random crystallites were generated dividing the layer plane into prescribed number of polygons created around randomly placed centers using the Voronoy-Delanay procedure. All crystallites had vertical boundaries (i.e., in the out-of-plane direction).

Two types of dislocations have been studied (separately). For the first type - vertical edge dislocations ($\mathbf{t} = \mathbf{e}_z$, $b_z = 0$), the in-plane components of their Burgers vector were chosen for each crystallite randomly and with equal probabilities from the set of six in-plane orientations $\mathbf{b} = \mathbf{a}_i$, possible in the hexagonal lattice. The length of the Burgers vector in this case was equal to the GaN lattice constant $a = 3.2$ Å. For mixed dislocations, the Burgers vectors for each grain were chosen randomly from the set $\mathbf{b} = \mathbf{a}_i + \mathbf{c}$, where $\mathbf{c} = \pm c \cdot \mathbf{e}_z$ with $c = 5.2$ Å [Mathis2001, Morkoc2008]. The same Burgers and direction vectors were assigned to all dislocations within the same grain.

Dislocation density $n_{dis}$ within each grain was assumed to be homogeneous in the layer plane. In the vertical direction, we have used for both studied dislocation types the density

$$\rho(z) = \frac{A_0}{z + z_0}, \qquad (19)$$

which rapidly decays with the distance $z$ from the film bottom. This simple functional form was suggested in [Mathis2001], basing on the analysis of available experimental data (see also [Bennett2010]). To determine the parameters $A$ and $z_0$, we have used the following two values of dislocation densities at two given heights: $\rho(z = 0.1\ \mu m) = 10^{10}$ cm$^{-2}$ and $\rho(z = 0.01$ mm$) = 10^9$ cm$^{-2}$.

Before proceeding to the discussion of simulation results, several comments concerning the relation between the proposed model and structure of real GaN films are in order. First, we concentrate ourselves on the stress induced by dislocations in the grain volume, postponing simulations and discussion of stress induced by grain boundaries (which can be viewed as dislocation walls) to the forthcoming publication. This means that the kind of polycrystallinity and types of grain boundaries occurring in real GaN layers are not subjects of this study. Next, the assumption that intergrain boundaries are vertical, represents a considerable simplification of the system, because in experiment one often observes the increase of the average GaN grain size with the film height, meaning that the coalescence of crystallites takes place during the film growth. Keeping in mind all these circumstances, we point out, however, that the thorough analysis of our simplified system is mandatory for understanding the properties of polycrystalline films with more complex structures.

## 6.1. Spatial distribution of the dislocation-induced stress: non-relaxed system

*Relation between the crystal grain structure and spatial stress pattern.* We start with the analysis of the relation between the polycrystalline layer structure and the spatial distribution of the dislocation-induced stress, obtained *before* the relaxation procedure for the proper consideration of elastic boundary conditions (described in Sec. 5 above). In Fig. 8 we first display the random grain structure used in simulations (a), directions of Burgers vectors for 'internal' grain (Burgers vectors in grains crossed by the border of the simulation area are not shown in order not to overload the picture) and (c) the difference between the Burgers vector in each discretization cell and the corresponding vectors in adjacent cells computed as $\Delta b = \frac{1}{4} \sum |\mathbf{b}_{ij} - \mathbf{b}_{i\pm1, j\pm1}|$. In our model this difference is zero for discretization cells *inside* each grain and can acquire a discrete set of non-zero values for cells adjacent to the grain boundaries (because the directions of Burgers vectors themselves have been chosen from a discrete sets of orientations, see above). Hence the brightness of each grain boundary visible in the panel (c) is proportional to the difference of the Burgers vectors of grains forming this particular boundary.

Comparison of the grain structure (images (a) - (c)) with the in-plane distribution of different stress tensor components shown in panels (d) - (f) (at the height $h = 2\ \mu$ above the layer bottom) immediately reveals strong correlations between the stress components and the grain structure. These correlations are qualitatively different for various components $\sigma_{\alpha\beta}$ and can be qualitatively understood using results presented above in Sec. 3 (Fig. 5).

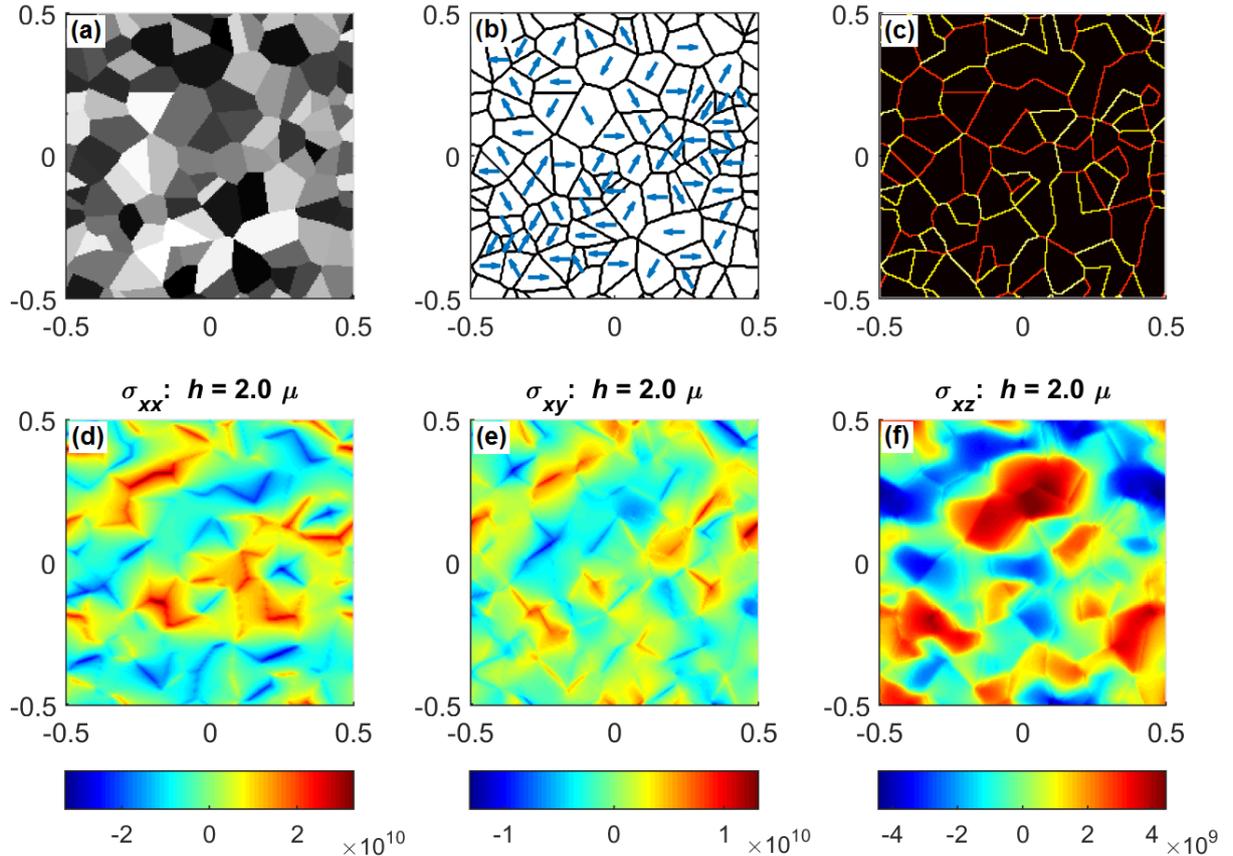

Fig. 8. Correlation between the random grain structure and the stress distribution in the GaN layer: (a) grain structure; (b) Burgers vectors of each grain, (c) difference between Burgers vectors in adjacent cells; (d) - (f) in-plane distribution of different stress tensor components at the height $h = 2\ \mu$ above the layer bottom. Lateral size of the simulated area $a_{sim} = 1.0$ mm, number of grains $N_{cr} = 100$ (resulting in the average grain size $d_{av} \approx 0.1$ mm), layer thickness $h = 0.5$ mm. Stress on colorbars is given in Pa.

For components $\sigma_{xx}$, $\sigma_{yy}$, $\sigma_{zz}$, and $\sigma_{xy}$, which are *odd* functions of in-plane coordinates $x$ and/or $y$ (see Eqs. (4)), contributions from dislocations inside the same grain are strongly 'averaged out', so that the magnitude of these components inside the grains are relatively low. On the other hand, near the grain boundary where Burgers vectors of adjacent grains have opposite directions, these components are strongly enhanced, because contributions from dislocations from adjacent grains near their mutual boundary sum up. Hence, the largest magnitude of arises at the grain boundaries. Second, the highest magnitudes of stress components $\sigma_{xx}$, $\sigma_{yy}$, $\sigma_{zz}$, and $\sigma_{xy}$ are observed on the boundaries between the grains $k$ and $l$ with the largest difference $|\Delta\mathbf{b}| = |\mathbf{b}_l - \mathbf{b}_k|$ between corresponding Burgers vectors. Examples of such patterns are clearly visible in Fig. 8 for $\sigma_{xx}$ and $\sigma_{xy}$ (panels (d) and (e)).

In contrast to this behaviour, off-diagonal components $\sigma_{xz}$ and $\sigma_{yz}$ are *even* functions of in-plane coordinates $x$ and $y$, so that contributions to these components from dislocations of the same grain accumulate within this grain. Hence these components reach their maximal magnitudes *inside* the grains - see panel (f) for $\sigma_{xz}$ in Fig. 8. According to Eqs. (4), the sign of $\sigma_{xz}$ and $\sigma_{yz}$ inside a particular grain depends respectively on the sign of $b_y$ or $b_x$ in this grain.

*Dependence of the stress distribution on the distance to the layer bottom.* Dislocations density (19) decays rapidly with increasing distance $z$ to the layer bottom. Hence the dislocations are mainly concentrated in this region, so that all $\sigma_{\alpha\beta}$ components have the maximal magnitude near the bottom.

Interestingly, different stress components demonstrate qualitatively different behaviour with increasing height ($z$-coordinate), what is due to two possible types of their $z$-dependencies present in expressions (4). Components $\sigma_{\alpha\beta}$ depend on the height $z$ above the layer bottom via the $R$ and $R_\lambda$ dependencies of corresponding expressions. Components $\sigma_{xx}$, $\sigma_{yy}$ and $\sigma_{xy}$ vary with $z$ only via the dependence $\sim 1/R_{\lambda+}^2$, and their lateral spatial distributions remain roughly the same. Only their magnitude decreases with increasing $z$ (and some in-plane smoothing also takes place), as shown on in-plane cuts in Fig. 9(a - c) and vertical cuts in Fig. 10(a,d) for the component $\sigma_{xx}$.

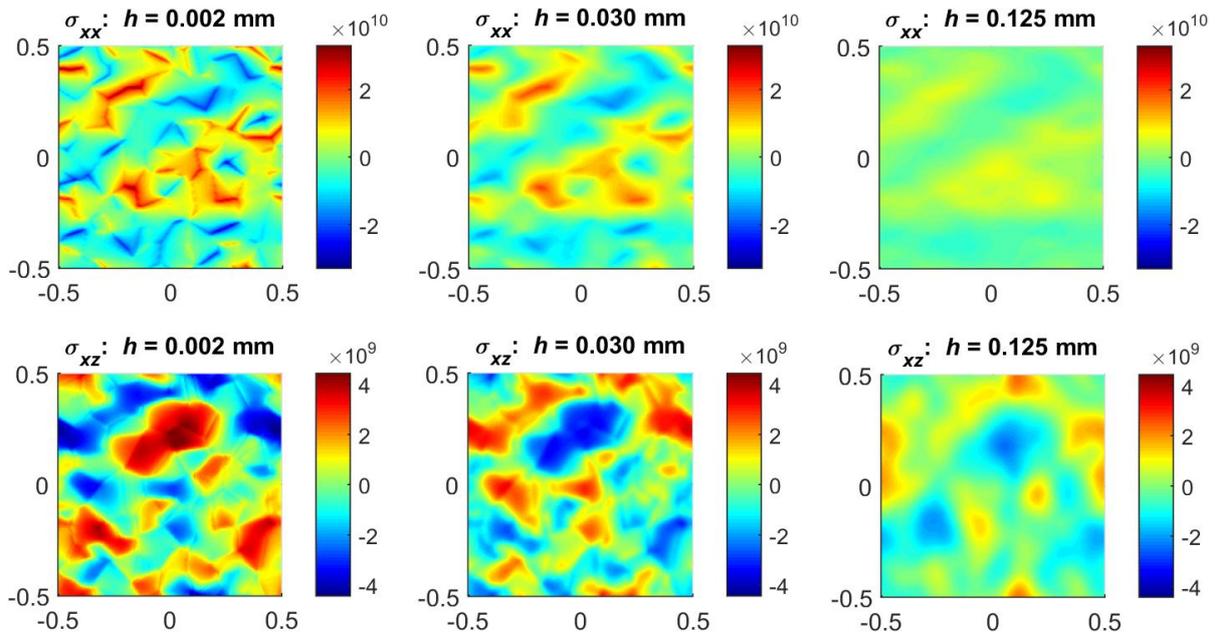

Fig. 9. Dependence of spatial patterns of $\sigma_{xx}$ (upper row) and $\sigma_{xz}$ (lower row) components on the height $h$ above the layer bottom for the same system as in Fig. 8. Stress on colorbars is given in Pa.

In contrast to this simple behaviour, components $\sigma_{zz}$, $\sigma_{xz}$ and $\sigma_{yz}$ change their sign with increasing the distance to the bottom plane, as shown on in-plane cuts in Fig. 9(d-f) for $\sigma_{xz}$ and on vertical cross-sections in Fig. 10(b,e) for $\sigma_{xz}$ and Fig. 10(c,f) for $\sigma_{zz}$. This sign change occurs due to the complicated $z$-dependencies of these components: they contain two terms with different power dependencies on $R$. E.g., the component $\sigma_{xz}$ is proportional to the expression $\sim (v/R - x^2/R^3)$, which changes its sign with increasing $z$ due to two $R$-dependencies having different powers and opposite signs. After the stress sign has been changed once, the in-plane spatial pattern of this second type of stress components remains the same, whereby their magnitude decreases analogous to the behaviour of $\sigma_{xx}$, $\sigma_{yy}$ and $\sigma_{xy}$.

To better visualize the qualitative picture explained above, we present in Fig. 10 typical vertical cross-sections of our layer (in the $yz$-plane), showing the evolution of the spatial distribution of both types of components. Note the different scaling of vertical axes in the upper and lower row of this figure. These two kinds of scaling were necessary to adequately present the evolution of $\sigma_{\alpha\beta}$ both across the whole layer thickness (linear $y$-axis, upper row) and the rapid decay of the stress magnitude and the sign change of some stress components near the layer bottom (logarithmic $y$-axis, lower row).

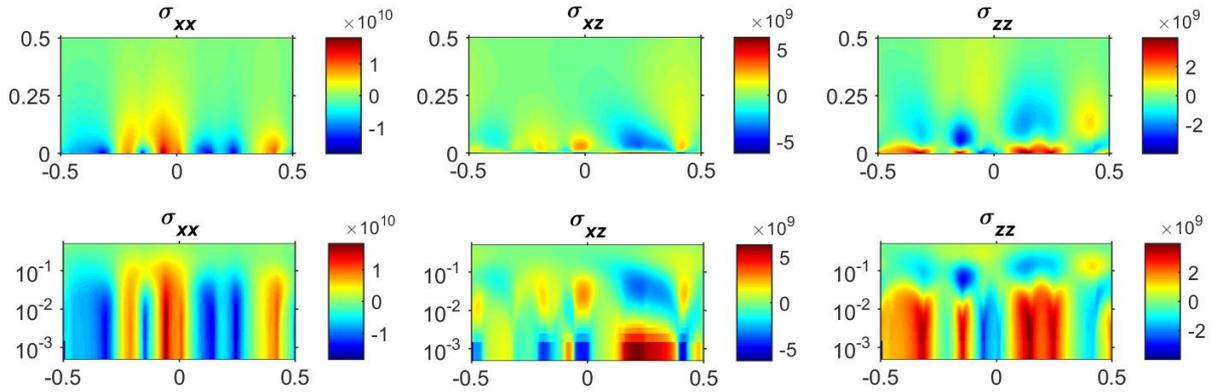

Fig. 10. Vertical ($yz$) cross-sections of the simulated layer, showing the patterns of $\sigma_{xx}$, $\sigma_{xz}$ and $\sigma_{zz}$ components using linear (upper row) and logarithmic (lower row) $z$-axes. The system is the same as in Fig. 8.

To quantify the dependence of the stress components on the distance to the layer bottom $h$, we show the in-plane standard deviations $s$ defined as

$$s_{\alpha\beta}^2(h_k) = \frac{1}{N_x N_y} \sum_{i,j} \sigma_{\alpha\beta}^2(x_i, y_j, h_k) \qquad (20)$$

of all components $\sigma_{\alpha\beta}$ as the functions of $h$ in Fig. 11. The definition (20) provides the mean-squared deviation of each stress components across the $xy$-plane at the height $h$ (we note that in our model the average values of $\langle \sigma_{\alpha\beta} \rangle = 0$ $\sigma_{\alpha\beta}$ themselves vanish due to the assumption of random directions of Burgers vectors). Hence $s_{\alpha\beta}$ can be considered as typical magnitudes of corresponding stress components at the height $h$. Rapid decay of all $s_{\alpha\beta}(h)$ is due to the following two features of our system: (*i*) the fast decrease of the average dislocation density (19) with increasing $h$ and (*ii*) alternating projections of Burgers vectors in different grains. The second feature leads to the fast decay of the typical stress in the same manner as the alternating charge density in an electrically charged line or plane leads to the fast (in some special cases - exponentially fast) decay of the electric field strength away from such systems.

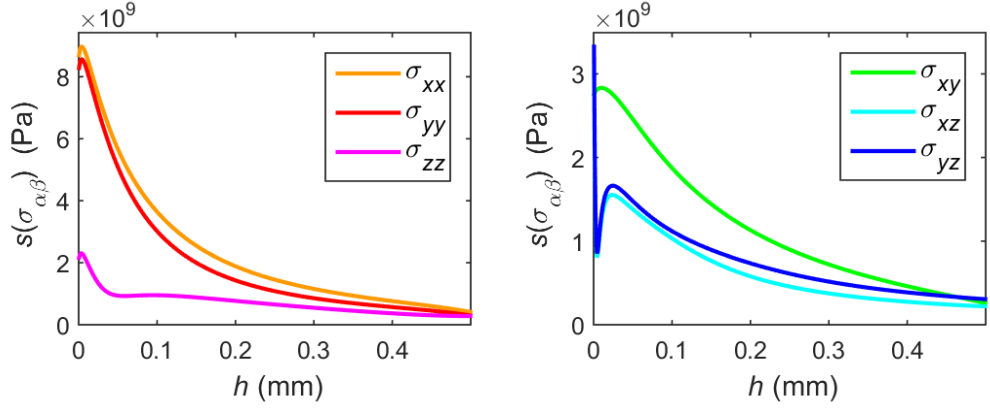

Fig. 11. In-plane standard deviation $s_{\alpha\beta}(h)$ as defined in Eq. (20) vs the height above the layer bottom $h$ for all stress tensor components.

The sharp minimum on $s_{xz}(h)$ and $s_{yz}(h)$ dependencies clearly visible (at very small $h$) on the panel (b) of Fig. 11 is due to the sign change of these components which was discussed above (see also maps of $\sigma_{xz}$ in Fig. 9 and 10): by changing sign, these components should pass through zero, so that their magnitude in this region is necessarily reduced.

We conclude this subsection with the following remark. In order to obtain the actual stress distribution in a free GaN, one should obviously apply the stress relaxation procedure described in Sec. 5 (see corresponding results in Sec. 6.2 below). However, results discussed in the current Section are relevant not only as the reference point for the analysis of the 'relaxed' stresses, but also may be close to the physical case where the GaN layer is still attached to a substrate, so that a kind of fixed boundary conditions for the bottom of GaN layer should be used.

### 6.2. Spatial distribution of the dislocation-induced stress: relaxed system

To obtain the spatial distribution of stress in a free (removed from the substrate) GaN layer, we have applied the relaxation procedure outlined in Sec. 5 to the system with the stress distribution discussed in the previous subsection. Free elastic boundary conditions (BC) were assumed on the top and bottom layer surfaces, whereas periodic BC were set on the side surfaces.

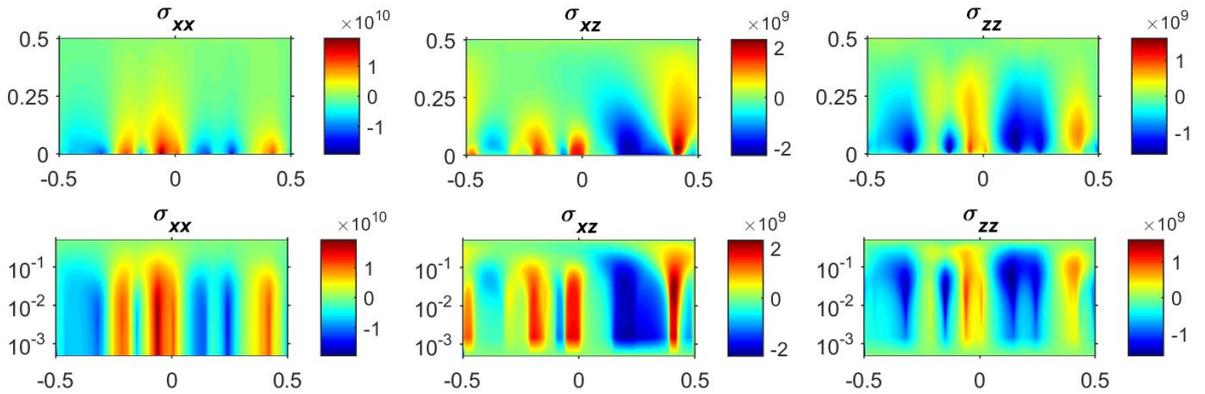

Fig. 12. Vertical cross-sections of our system, showing the patterns of $\sigma_{xx}$, $\sigma_{xz}$ and $\sigma_{zz}$ in the same way (linear and logarithmic z-axes) as in Fig. 10, but *after* application of the relaxation procedure described in Sec. 5. Note qualitative changes in the spatial distribution of $\sigma_{xz}$ and $\sigma_{zz}$ components as compared to Fig. 10.

Results of this simulation - spatial distribution of the relaxed stress components and height dependence of standard deviations $s_{\alpha\beta}(h)$ - are shown in Fig. 12 and Fig. 13, correspondingly. First of all, spatial distribution and values of those stress components which do not contain the $z$-index ($\sigma_{xx}$, $\sigma_{yy}$ and $\sigma_{xy}$) change relatively weakly: these components are not directly affected by the free BCs (16b), so they change only because various stress components are 'connected' via the (relatively small) Poisson ratio $\nu$.

In contrast to this behaviour, both the distribution and the values of '$z$-containing' components $\sigma_{zz}$, $\sigma_{xz}$ and $\sigma_{yz}$ changes qualitatively, because on the top and bottom surfaces these components have to obey the free BCs (16b). Especially the BC on the bottom surface, where the initial (non-relaxed) stress is at its maximum, play a very important role. Comparison of non-relaxed (Fig. 10) and relaxed (Fig. 12) distributions of $\sigma_{xz}$ and $\sigma_{zz}$ demonstrates that BCs forcing zero values of these components on the bottom surface result, first, in the large decrease of their maximal values - see corresponding color bars - and to the sign change of the dominant contribution for the given in-plane location - compare vertical stress 'bands' on corresponding figures with the log-scaled $z$-axes.

Qualitatively the same conclusions can be drawn from the comparison of the standard deviations for non-relaxed (solid lines) and relaxed (dashed lines) stresses (Fig. 13), where the dependencies $s_{\alpha\beta}(h)$ for $\sigma_{xx}$, $\sigma_{yy}$ and $\sigma_{xy}$ differ only slightly, whereas $s_{\alpha\beta}(h)$ for $\sigma_{zz}$, $\sigma_{xz}$ and $\sigma_{yz}$ undergo qualitative changes.

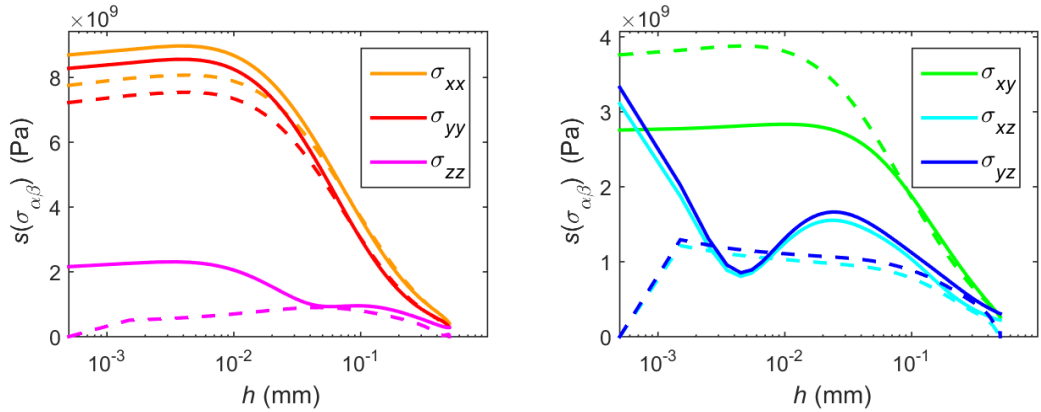

Fig. 13. Comparison of in-plane standard deviations $s_{\alpha\beta}(h)$ defined via (20) for all stress tensor components before (solid lines) and after (dashed lines) the application of the relaxation procedure. Note the log-scale of the horizontal axis ($h$-axis).

We also note that the distance to the layer bottom, after which the stress components begin to decrease rapidly ('localization height' of the stresses), is approximately equal to the lateral grain size, in this case $\approx 0.1$ mm. This feature confirms that the overall fast decrease of stress in this system is indeed due to the alteration of Burgers vectors projections in different grains, as explained above.

A more detailed information concerning the stress distribution density can be extracted from histograms of the stress distribution at a specific height from the layer bottom.

As an example, in Fig. 14 we present the evolution of such histograms for the $\sigma_{zz}$-component with increasing height $h$ before (upper row in the figure) and after (lower row) the stress relaxation has been performed. The strong influence of the stress relaxation both on the histogram width and on the evolution of the stress distribution with varying height for this particular stress component can be clearly recognized. In contrast to $\sigma_{zz}$, the distribution and the evolution character of other diagonal components with $h$ change only insignificantly after the stress relaxation (for the same reason as corresponding standard deviations - these components are much less affected by the elastic boundary conditions at the top and bottom layer surfaces).

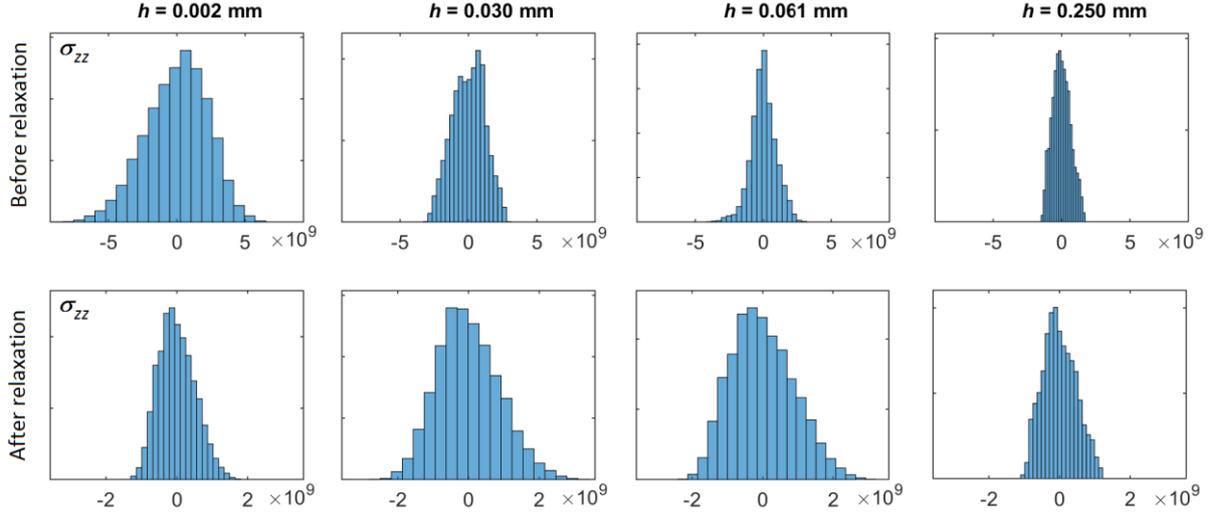

Fig. 14. Histograms of the distribution of $\sigma_{zz}$ stress components in horizontal ($xy$) planes at different heights from the layer bottom before (upper row) and after (lower row) the relaxation procedure.

### 6.3 Stress for the system of 'mixed' dislocations

In order to compare the stress patterns generated by various dislocation types, we have simulated also a system with the height-dependent dislocation density given by (19), but with 'mixed' dislocations. As noted at the beginning of Sec. 6, Burgers vectors of these dislocations have the form $\mathbf{b} = \mathbf{a}_i + \mathbf{c}$, where $\mathbf{c} = \pm c \cdot \mathbf{e}_z$, where vectors $\mathbf{a}_i$ (6 possible orientations) and the sign before the $c$-component have been chosen randomly for each crystallite.

Results of this simulation are shown in Fig. 15 as vertical cross-sections of spatial distributions for stress components $\sigma_{xx}$, $\sigma_{xz}$ and $\sigma_{zz}$, before and after the stress relaxation procedure; note the logarithmic scale of the vertical spatial axis. We have used the same random polycrystalline structure as for simulation of vertical edge dislocations, in order to enable a quantitative comparison of both cases. This comparison (see the lower image rows in Fig. 10 and 12) reveals that spatial patterns of stress components $\sigma_{xx}$ and $\sigma_{zz}$ for by these two dislocation types are very similar. However, the magnitude of the non-diagonal component $\sigma_{xz}$ is much larger for mixed dislocations and the spatial distribution of $\sigma_{xz}$ is qualitatively different from the edge dislocation case; the same applies to $\sigma_{yz}$ (results not shown).

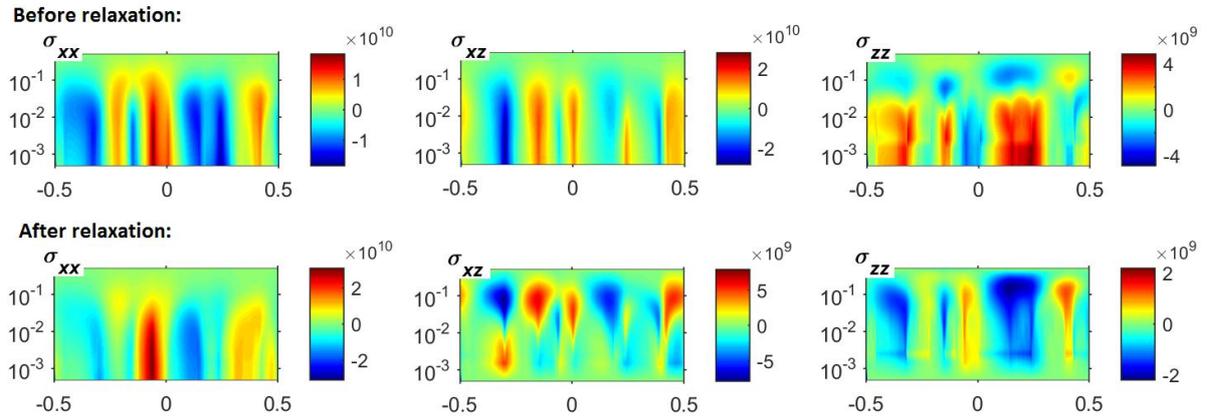

Fig. 15. Vertical cross-sections of the distribution of stress components $\sigma_{xx}$, $\sigma_{xz}$ and $\sigma_{zz}$ in a system of *mixed* dislocations, shown using the logarithmic $z$-axes before (upper row) and after (lower row) application of the relaxation procedure.

This behaviour can be explained qualitatively when we neglect the relatively small incline ($\alpha = 15.6^{\circ}$) of mixed dislocations in GaN with respect to the vertical axis and consider a mixed dislocation in the first approximation as a 'superposition' of a vertical edge dislocation with the corresponding Burgers vector $\mathbf{b} = \mathbf{a}_i$ and a vertical screw dislocation with $\mathbf{b} = \pm\mathbf{c}$. In this model the relatively simple analytical expressions (4) - instead of the practically untreatable formulas (6) and (7) can be used for the stress analysis. This analysis immediately shows that the screw dislocation part, being the only part with the non-zero $b_z = \pm c$, does not contribute to diagonal stress components. This feature explains very similar distributions of $\sigma_{xx}$ and $\sigma_{zz}$ for vertical edge and mixed dislocation before the relaxation procedure was applied.

In contrast, the non-diagonal components $\sigma_{xz}$ and $\sigma_{yz}$ contain a significant contribution from the screw dislocation part with $b_z \neq 0$ - see the last two formulas in (4). The contribution to $\sigma_{xz}$ coming from the term with $b_z$ is expected to be even larger than those from terms proportional to $b_x$ and $b_y$, because (*i*) the in-plane self-averaging of the $b_z$-contribution (~ $y$) is not as strong as of the $b_x$-term (~ $x \cdot y$), (*ii*) it does not change sign in contrast to the $b_y$-term and (*iii*) the magnitude of the $b_z$-component is larger than of $b_x$ and $b_y$ ($c > a$). As the result of these factors, the screw component makes a dominant contribution to $\sigma_{xz}$ and $\sigma_{yz}$, which does not change sign in the vertical direction and achieves much larger maximal values than for vertical edge dislocations ($\approx 30$ GPa vs $\approx 6$ GPa). This naturally leads to a very different spatial patterns also after the stress relaxation have been performed.

### 6.4. Dependence of the stress distribution on the average crystal grain size: simulation results

One of the most important questions in applications of the dislocation theory is the dependence of the dislocation-induced stress on the average grain size in polycrystalline films. In this subsection we address this question using numerical simulations, and in the next subsection we present analytical predictions of this dependence for two practically important cases.

In order to study this problem using computer modeling, we have simulated systems with the same height $h = 0.5$ mm and four different average grain sizes in the layer plane: $D_{cr} = 0.001$, 0.01, 0.1 and 1.0 mm. In order to compare not only statistically averaged characteristics of stress, but also its spatial patterns for different $D_{cr}$, we have used the same random grain arrangement (shown in Fig. 8) in all simulation runs, scaling this particular grain pattern to corresponding average grain sizes. All systems have been simulated using periodic BC in the layer plane, where the simulated region represents a square with the side $L = 10 \cdot D_{cr}$. By choosing the number of discretization cells in each spatial direction, we have aimed to reach the best compromise between the requirements (*i*) to discretize each grain sufficiently fine in the layer plane (to resolve the stress inside each grain), (*ii*) to obtain the adequate discretization perpendicular to the layer plane (because the dislocation density rapidly decreases with $h$, see Eq. (19)) and (*iii*) to keep the relation between the in-plane and out-of-plane sizes of a single discretization cell in reasonable limits (otherwise the computational time for the procedure evaluating the interaction coefficients between non-cubic cells described in subsection 2.3 would become too long). These consideration have led to discretization cell numbers $L_x \times L_z = 120 \times 150$, $200 \times 500$, $200 \times 500$ and $400 \times 200$ for the four values of $D_{cr}$ listed above, correspondingly (note that $L_y = L_x$). Results shown below are obtained without the stress relaxation, in order to enable the comparison with analytical theory in subsection 6.5.

Selected simulation results are shown in Fig. 16 (spatial patterns of the stress component $\sigma_{xx}$) and 17 (height dependencies of the standard deviation for $\sigma_{xx}$, $\sigma_{xy}$ and $\sigma_{zz}$). Spatial distribution of the $\sigma_{xx}$ stress component demonstrates typical trends for the stress pattern when the average grain size is increased (columns from left to right in Fig. 16): (*a*) the maximal and average stress values increase and (*b*) the stress distribution, which is strongly non-homogeneous with maxima along the grain boundaries near the layer bottom (upper row in this figure), becomes smoothed when the height (distance to the layer bottom) increases, and this smoothing at the given height is stronger for smaller grain sizes - compare e.g., 2$^{nd}$ and 3$^{rd}$ columns in Fig. 16.

The first trend - decrease of maximal stress values with decreasing grain size - can be explained by the contribution from neighboring grains to the stress observed at any point of the system. Namely, for large grain sizes maximal stress is concentrated at the grain boundaries with the largest 'jumps' of the Burgers vectors as explained in subsection 6.1. When the grain size decreases, contributions coming from neighboring grains become more significant (distance to neighboring grains decreases). As this contribution is random, in most cases it will reduce maximal stress values, concentrated at these locations (see first row in Fig. 16).

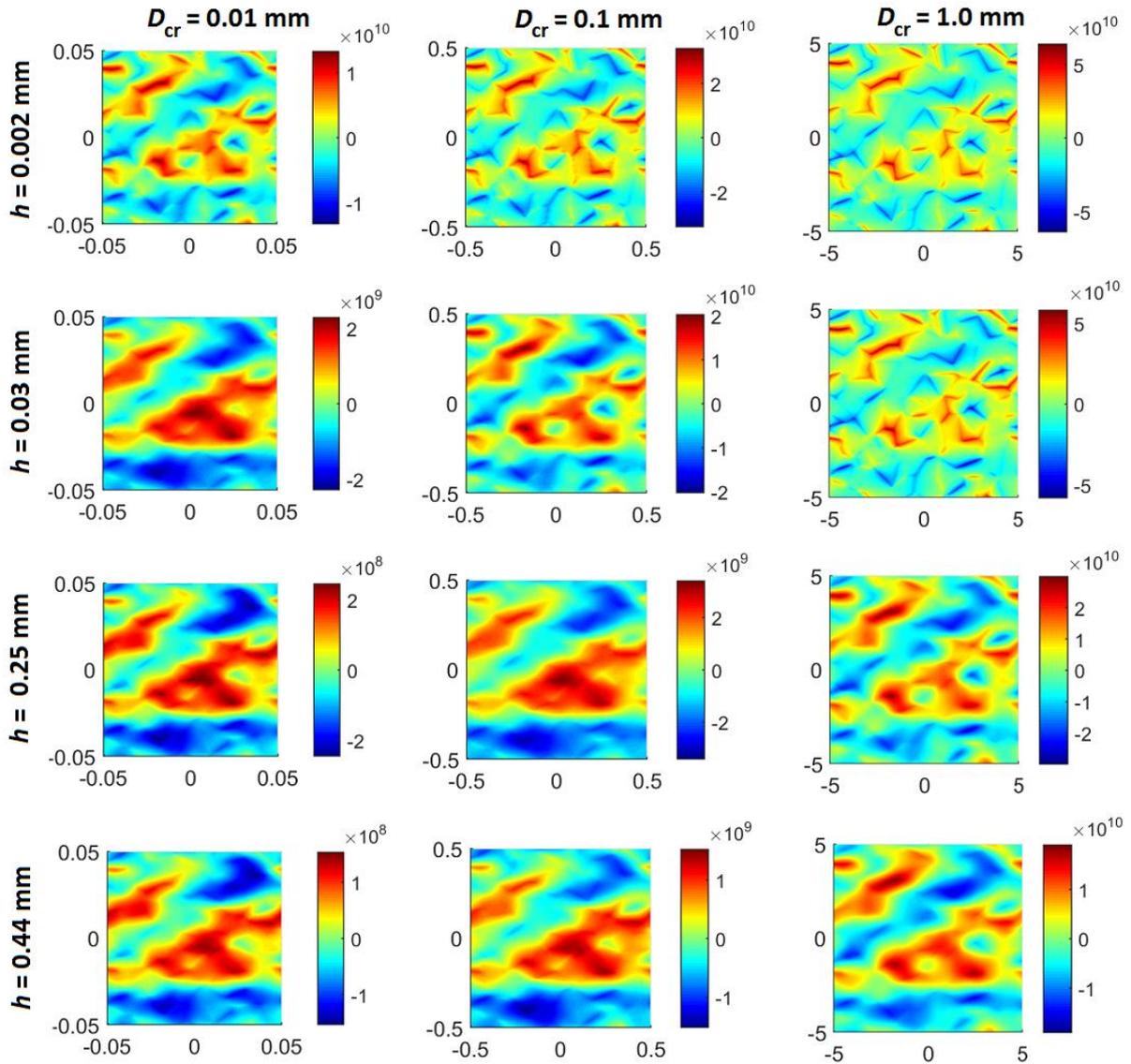

Fig. 16. Change of the spatial distributions of the $\sigma_{xx}$-component in a system of vertical dislocations with random distributions of Burgers vectors among crystal grains, when the average grain size (different columns) and heights from the layer bottom (different rows) are varied.

The second trend - faster decay and stronger smoothing of the stress for smaller grains, when the distance $h$ to the layer bottom increases - is due to the rapidly decreasing dislocation density and the effect of the random sign of Burgers vector projections in different grains. As explained in subsection 6.1 - see discussion of Fig. 11 after the Eq. (20) - these alternating signs lead to the rapid decay of the total stress when the distance to the layer bottom (where most dislocations are concentrated) increases. The characteristic height of this decay approxi-mately corresponds to the lateral grain size, as it is always the case for such random systems. This trend is also clearly seen in Fig. 17, especially for $\sigma_{xx}$ and $\sigma_{xy}$ components.

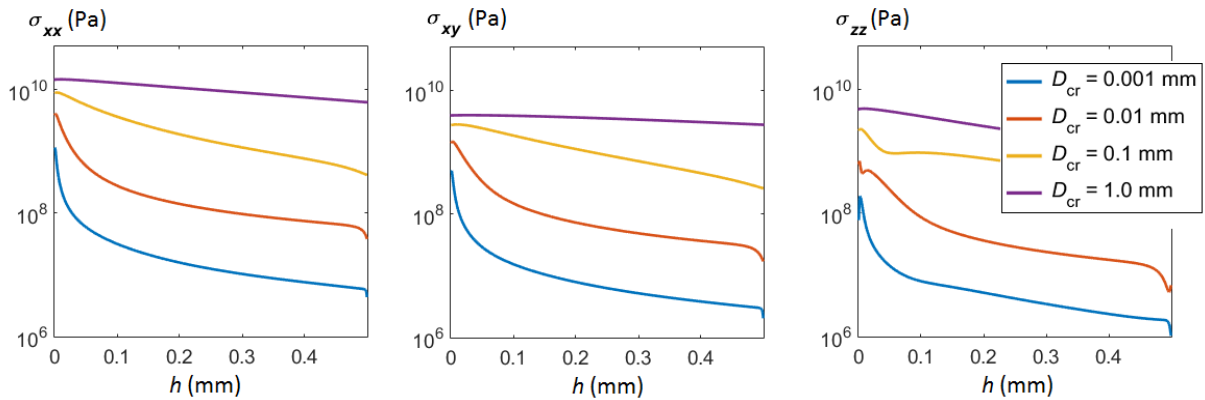

Fig. 17. In-plane standard deviation $s(h)$ as defined in Eq. (20) vs the height above the layer bottom $h$ for different average grain sizes $D_{cr}$ as shown in the legend (for components $\sigma_{xx}$, $\sigma_{xy}$ and $\sigma_{zz}$).

### 6.5. Stress dependence on the average crystal grain size: analytical study

In this Section we develop a general method for the analytical estimation of the standard deviations of stress components in a polycrystalline film in dependence on the average lateral grain size $D_{av}$. For a height-dependent dislocation density analytical solution of this task is not really possible, so in this part we consider a system with vertically homogeneous (height-independent) density. Results obtained in this subsection allow several important insights in the physics of dislocation systems and are also highly instructive for the understanding of our simulation results (although direct comparison with simulation data presented in previous section cannot be performed).

In a polycrystalline film, the total stress at any given point is the sum of contributions coming from dislocations in all grains. Thus in a film with a random grain structure the stress components at a given point $\sigma_{\alpha\beta}(\mathbf{r})$ can be considered as independent random variables. According to the central limit theorem (see, e.g., [Shiryaev1996]) the distribution density of $\sigma_{\alpha\beta}(\mathbf{r})$ should in this case be approximately Gaussian, and the variance $s_{tot}^2$ of the total stress at some point can be computed as a sum of variances $s_i^2$ of partial stresses induced at this point by dislocations from all grains $i$ of the film.

In the simplest approximation (Fig. 18) the grains can be considered as being arranged in spherical shells. In this model, for a system with the average grain diameter $D_{av}$ we can easily estimate the number of grains containing in each shell and the distance between the grains and the target point.

From the structure shown in Fig. 18, we can deduce that the distance between the centers of grains belonging to the first shell and the center of the target grain is $\Delta r_1 \approx D_{av}$, and the number of grains in this shell is $n_1 \approx 6$. Similarly, for the $k$-th shell the distance to the target grain is $\Delta r_k \approx k \cdot D_{av}$, and the number of grains is $n_k \approx l_k / D_{av} \approx 2\pi \cdot \Delta r_k / D_{av} \approx 2\pi k$.

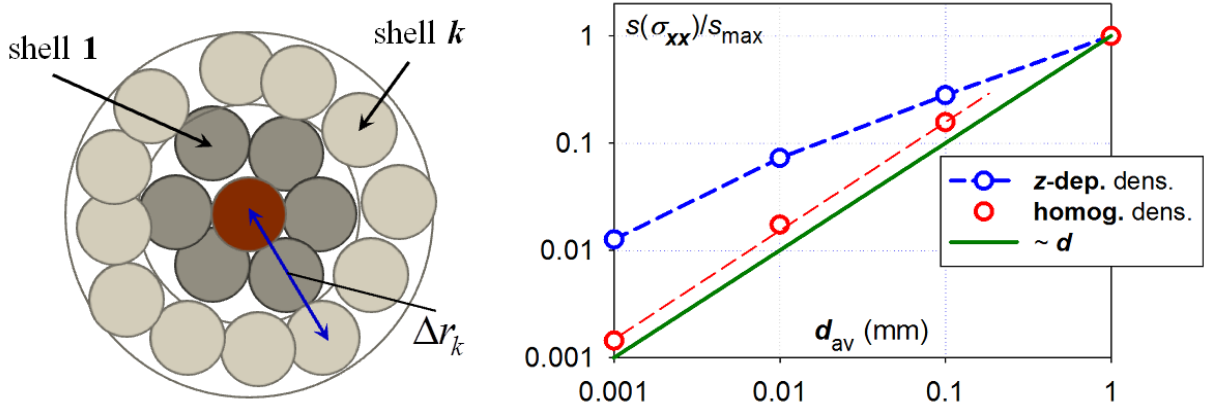

Fig. 18. **Left** panel: construction of the shells surrounding the center target shell (dark red) for the analytical estimation of the variance of the dislocation-induced stress in a polycrystalline film. **Right** panel: comparison of predictions of the analytical theory (green solid line) with numerical results for height-dependent dislocation density (blue open circles, blue dashed line is the guide for an eye) and height-independent density (red open circles). Red dashed line in the latter case has the same slope as the analytical prediction for this system.

As the total stress is the sum of contributions from all shells, the variance of this total stress can be estimated employing the central limit theorem in the following manner:

$$s_{tot}^2 = \sum_{k=1}^{\infty}\sum_{i=1}^{n_k} s_i^2(k) \approx 2\pi \sum_{k=1}^{\infty} k \cdot s_1^2(k) \quad (21)$$

Here we have assumed that the variance of stresses from all grains belonging to the same shell is approximately the same and used the result for the number of grains in the $k$-th shell $n_k \approx 2\pi k$ derived above.

The stress generated on the target grain by a grain $i_k$ from the $k$-th shell is proportional to the dislocation density $\rho_d$ (we assume that this density is approximately the same for all grains) and the area of the source grain $S_i$. Hence we can estimate this stress as

$$\sigma_{\alpha\beta}(i_k) = a_{\alpha\beta}(\varphi) \cdot \frac{\rho_d \cdot S_i}{(\Delta r_k)^m} \quad (22)$$

here the function $a_{\alpha\beta}(\varphi)$ reflects the dependence of the stress component $\sigma_{\alpha\beta}$ on the angle $\varphi$ between the $x$-axis of the in-plane coordinate system and the radius-vector pointing towards the grain $i_k$. The stress decays with the distance $\Delta r_k$ between the target grain and the source grain from the shell $k$ as $\sim 1/(\Delta r_k)^m$; the value for the decay power $m$ will be discussed below. Taking into account that the distances $\Delta r_k$ are the same for all grains from the shell $k$ and the angle $\varphi$ varies in the limits $0 \leq \varphi < 2\pi$, we obtain the variance of the stress from one grain of the $k$-th shell using (22) as

$$s_1^2\left(\sigma_{\alpha\beta}^{(k)}\right) = \frac{1}{2\pi}\int_\theta \sigma_{\alpha\beta}^2(i_k,\varphi)\cdot d\varphi = \left(\frac{\rho_d \cdot S_1}{(\Delta r_k)^m}\right)^2 \cdot \frac{1}{2\pi}\int_0^{2\pi} a_{\alpha\beta}^2(\varphi)\cdot d\varphi = g_{\alpha\beta}\left(\frac{\rho_d \cdot S_1}{(\Delta r_k)^m}\right)^2 \quad (23)$$

where we have introduced the notation $g_{\alpha\beta} = \langle a_{\alpha\beta}^2 \rangle$.

Substituting the average grain area $S_{av} = \pi D_{av}^2/4$ and the distance $\Delta r_k \approx k \cdot D_{av}$ into (23), for the variance of the stress induced by one grain from the shell $k$ we obtain:

$$s_1^2\left(\sigma_{\alpha\beta}^{(k)}\right) \approx \frac{\pi^2}{16} g_{\alpha\beta} \cdot \rho_d^2 \cdot \frac{D_{av}^4}{(k \cdot d_{av})^{2m}} \quad (24)$$

Using the number of grains in the $k$-th shell $n_k \approx 2\pi k$, and summing contributions from all shells up to the maximal shell number $k_{max}$ (see additional discussion below), we finally obtain the following dependence of the variance $s_{tot}^2$ for the total stress components on the average grain size:

$$s_{tot}^2 = \sum_{k=1}^{k_{max}} s_i^2\left(\sigma_{\alpha\beta}^{(k)}\right) \cdot 2\pi k = \frac{\pi^3}{8} \cdot g_{\alpha\beta} \cdot \rho_d^2 \cdot \sum_{k=1}^{k_{max}} \frac{1}{k^{2m-1}} \cdot D_{av}^{4-2m} \qquad (25)$$

The general result (25) allows us to consider limiting cases of 'flat' ($D_{av} \gg h$) and 'high' ($D_{av} \ll h$) grains separately. For 'flat' grains, i.e. for layers with the thickness much smaller than the grain size, the dislocation-induced stress decays with distance $r$ between the dislocation segment and the target point as $\sim 1/r^2$ (see the discussion in Sec. 4 after Fig. 7), so that the decay power in (25) is $m = 2$. Hence the variance $s_{tot}^2$ (and the standard deviation $s_{tot}$ itself)

$$s_{tot}^2 \sim \sum_{k=1}^{k_{max}} \frac{1}{k^{2m-1}} \cdot D_{av}^{4-2m} = \sum_{k=1}^{k_{max}} \frac{1}{k^3} \neq f(D_{av}) \qquad (26)$$

does not depend on the grain size.

The physical reason for this apparently counterintuitive result is the following. Typical stress produced by a source grain, is proportional to its area (see (22)), so that $\sigma_{\alpha\beta} \sim D_{av}^2$. On the other hand, this stress is inversely proportional to the $m$-th power of the distance $\Delta r_k$ between source and target grains: $\sigma_{\alpha\beta} \sim (\Delta r_k)^{-m}$. For the structure shown in Fig. 18, this distance is proportional to the average grain size: $\Delta r_k \approx k \cdot D_{av}$. Hence for 'flat' grains ($m = 2$) the stress increase due to the growth of the source grain area is exactly compensated by the stress decrease due to larger intergrain distances in a system with larger grains. We note that the sum over $k$ in (26) rapidly converges with increasing $k_{max}$, so that we can use $k_{max} \to \infty$.

In the opposite case of $h \gg D_{av}$, i.e. for films with the thickness much larger than the average grain diameter, the distance between the source and target grains is much smaller than the dislocation length (at least for several nearest neighboring shells). Hence for this geometry the dislocation-induced stress decays with $r$ as in the 2D case of an infinitely long dislocation, namely as $\sim 1/r$. This means that the decay power in (22) is $m = 1$, resulting to the variance dependence on the grain size in the form

$$s_{tot}^2 \sim \sum_{k=1}^{k_{max}} \frac{1}{k^{2m-1}} \cdot D_{av}^{4-2m} = D_{av}^2 \cdot \sum_{k=1}^{k_{max}} \frac{1}{k}, \qquad (27)$$

Hence for this system the standard deviation of the total stress is proportional to the grain size: $s_{tot} \sim D_{av}$. The physical reason for this dependence is that for $h \gg D_{av}$ the stress increases with the source grain size faster ($\sigma_{\alpha\beta} \sim D_{av}^2$), than it decreases due to the increase of the intergrain distances ($\sigma_{\alpha\beta} \sim 1/\Delta r_k \sim 1/D_{av}$).

Regarding the expression (27), we note that from the formal point of view, the sum over $k$ diverges weakly (logarithmically) with growing $k_{max}$. In a real physical system, the convergence is warranted either by the finite lateral size of the film (for very small samples) or when the shell radius $\Delta r_k$ becomes larger than film thickness.

Concluding this discussion, we would like to point out, that both power dependencies used by the derivation of (26) and (27), namely $\sigma_{\alpha\beta} \sim D_{av}^2$ and $\Delta r_k \sim D_{av}$, are not specific for our shell model, but rather universal for a polycrystalline film. In particular, the dependence $\sigma_{\alpha\beta} \sim D_{av}^2$ is a simple consequence of the fact that in the first approximation the stress induced by all dislocations from a source grain on a target grain can be considered as a stress produced by a 'macrodislocation'. This 'macrodislocation' is located in the center of the source grain and has the Burgers vector equal to the sum of all Burgers vectors of dislocations inside a source grain. The sum over *k* (with the proportionality $\Delta r_k \sim D_{av}$) is specific for this shell model, and should in principle be replaced in a rigorous model of a random polycrystalline structure by the integral over the intergrain distances $\Delta r$ with the weighting function $f(\Delta r)$ showing how many grains are located on average at the distance $\Delta r$ from the target grain. It is straightforward to show that this replacement would preserve the power dependencies obtained in (26) and (27), although prefactors in these dependencies will be somewhat different from those obtained in our shell model.

In order to directly compare our analytical predictions with numerical results, we have simulated a layer with vertically homogeneous dislocation density and the thickness *h* = 2 mm. Simulated dependence of the standard deviation $s_{tot}^{xx}$ for the $\sigma_{xx}$-component on the grain size $D_{av}$ is shown in Fig. 18 on the right panel with red open circles. Results are normalized to the maximal *s*-value achieved for the largest $D_{av}$. The slope of this dependence coincides very well with the analytically predicted law (27) $s_{tot} \sim D_{av}$ (shown with the solid green line) up to the grain size $D_{av}$ = 0.1 mm (dashed red line); the value of $s_{tot}$ for $D_{av}$ = 1 mm does not lie on this line anymore, because for this size the condition $h \gg D_{av}$ is not fulfilled.

For the layer with the *z*-dependent dislocation density (19) the majority of dislocations is concentrated near the bottom of the film, so that the 'effective' grain thickness $h_{eff}$ is much lower than the actual film thickness *h*. If this 'effective' thickness would be always much smaller than the lateral grain size $d_{av}$, then standard deviation $s_{tot}$ would be independent on $D_{av}$, according to (26). In our case we have an intermediate situation, so that $s_{tot}^{xx}$ still increases with increasing grain size, but not as fast as for a system with the vertically homogeneous dislocation density (blue circles connected with the dashed line in Fig. 18).

### 6.6 Comparison to experimental results

Although the importance of the dislocation-induced stress in general and in GaN layers in particular is of a large importance, only a few experimental results on this topic are available. The main reason for this deficiency is the difficulty of corresponding experimental measurements, where both the site-dependent stress pattern and dislocation density have to be explored for one and the same sample or at least for GaN layers belonging to the same charge obtained under identical growth conditions.

Due to this problem, only a qualitative comparison of our results with the measurement data is possible. In particular, in one of the earlier papers [Hearne1999], the tensile stress in the range of 0.1 to 0.3 GPa for MOCVD-grown GaN films with the thickness of several $\mu$m was measured. This thickness corresponds to the initial parts of our simulated dependencies $\sigma_{\alpha\beta}(h)$. Thus the values found in [Hearne1999] are in a good qualitative agreement with our values for small *h*. However, the origin of the stress measured in [Hearne1999] was not clarified; the authors could exclude the pseudomorphic growth of GaN on a thermally strained buffer and the onset of the island coalescence during the growth as reasons of the observed tensile stress, leaving crystal defects as one of possible stress origins.

In the work of Faleev et al. [Falleev2005] the elastic stress in the range 0.1 - 1.3 GPa was observed in HVPE-grown GaN films on SiC substrate for film thicknesses between 0.3 and 30 $\mu$m. This value is also in a qualitative agreement with our results. Authors of [Falleev2005] have tried to correlate the measured stress with the dislocation density; however, only screw dislocations have been taken into account in [Falleev2005], so that the establishing of a quantitative relation to our data is unfortunately not really possible.

In the recent publication [Barchuk2014] correlations between the residual stress and the density of threading dislocations also in HVPE-grown GaN films were studied. It was found that the total dislocation density decreases with the increasing film thickness and that the stress measured on the film surface also decreases roughly from ≈ 0.5 GPa for very thin films ($h \sim 10~\mu$m) to ≈ 0.05 GPa for films with the thickness of $h \approx 900~\mu$m. This results also agree fairly well with our findings, but a more detailed sample characterization and the site-dependent measurements of both the stress and the dislocation density are necessary for the quantitative comparison between simulations and experiment.

## 7. Conclusion

In this study we have developed a fast and accurate method for the computation of stress generated in crystal layers by dislocations with arbitrary directions and Burgers vectors. The method is based on the dislocation density formalism and employs the FFT for the evaluation of stress components. Further, it allows to use discretization cells with arbitrary shape anisotropy, thus strongly reducing computational time for systems, where the spatial variation of the dislocation density requires significantly different mesh sizes along different spatial directions.

Using this method, we have studied in detail the dislocation-induced stress in a polycrystalline layer with a random grain structure, containing dislocations of various types (vertical edge and mixed dislocations). We have considered a layer made of GaN, because this material is widely used in various applications. Further, we have assumed that the dislocation density rapidly varies in the vertical direction ($z$-direction), as it usually the case for HVPE-and MOVPE-grown GaN layers. Under this assumption we observe a strong variation of typical stress values with the distance to the layer bottom. We have also shown, that for a layer containing vertical dislocations, diagonal and non-diagonal stress components exhibit qualitatively different behaviour. In addition, stress components $\sigma_{\alpha z}$ containing the spatial index $z$ (corresponding to the direction perpendicular to the layer plane), behave also very different compared to other stress components. All these features could be explained basing on the analytical expressions for the stress induced by a single vertical dislocation. For a layer containing mixed dislocations, we have found that whereas diagonal stress components are similar to the system with vertical dislocations only, non-diagonal components (especially $\sigma_{xz}$ and $\sigma_{yz}$) possess very different features due to the presence of a strong screw part in these dislocations.

Finally, we have investigated the dependence of the stress pattern and typical stress values on the average grain size $d_{av}$ of the layer. We have demonstrated that both the typical stress values and the characteristic decay length of the stress in the $z$-direction increase with the average grain size. In addition, we have succeeded to develop an analytical model which predicts the stress dependence on $d_{av}$ for a system with vertically homogenous dislocation density and presented the comparison of our simulation results with analytical predictions for various systems.

**ACKNOWLEDGEMENTS**. We greatly acknowledge many fruitful discussions with Dr. B. Weinert, Dr. M. Jurisch, Dr. S. Eichler and other colleagues from Freiberger Compound Materials GmbH (FCM), and Prof. P. Görnert. We also acknowledge financial support of FCM, Saxony Ministry of Economics (projects 70890/1949 and 100201540/2955) and German Federal Ministry of Education and Research (project 16BM1200).